\documentclass[thmsa,notitlepage, 12pt,letterpaper]{article}
\usepackage{amssymb}
\usepackage{amsfonts}
\usepackage{amsmath}
\usepackage{harvard}
\usepackage[textwidth=6in, textheight=8.5in]{geometry}
\usepackage[nodisplayskipstretch]{setspace}

\input{tcilatex}
\begin{document}

\title{\textbf{{\Large Quantile Dependence between \textbf{Stock Markets and 
}Its Application in Volatility Forecasting\thanks{%
I would like to thank Robert F. Engle, Simone Menganelli and the seminar
participants at Sungkyunkwan University, the 9th SoFiE (Society of Financial
Econometrics) annual conference (Hong Kong), 2016 Time Series Workshop on
Macro and Financial Economics (Seoul), the 10th Cross-Strait Conference on
Statistics and Probability (Chengdu), and 2016 Korean Econometric Society
Summer Meeting (Jeju) for their valuable comments and suggestions.}}}}
\author{Heejoon Han\thanks{%
Department of Economics, Sungkyunkwan University (E-mail:
heejoonhan@skku.edu). }}
\date{August 2016}
\maketitle

\begin{abstract}
\onehalfspacing This paper examines quantile\ dependence between
international stock markets and evaluates its use for improving volatility
forecasting. First, we analyze quantile dependence and directional
predictability between the US\ stock market and stock markets in the UK,
Germany, France and Japan. We use the cross-quantilogram, which is a
correlation statistic of quantile hit processes. The detailed dependence
between stock markets depends on specific quantile ranges and this
dependence is generally asymmetric; the negative spillover effect is
stronger than the positive spillover effect and there exists strong
directional predictability from the US market to the UK, Germany, France and
Japan markets. Second, we consider a simple quantile-augmented volatility
model that accommodates the quantile dependence and directional
predictability between the US market and these other markets. The
quantile-augmented volatility model provides superior in-sample and
out-of-sample volatility forecasts.
\end{abstract}

\noindent \textit{Keywords:} Quantile, Cross-quantilogram, Spillover,
Volatility Forecasting.

\doublespacing

\section{Introduction}

In many circumstances, investors are interested in dependence between
financial markets such as dependence between international stock markets,
dependence between currency markets, dependence between stock markets and
bond markets or dependence between stock markets and commodity markets. It
is essential for investors to have an understanding of the dependence
between financial markets because this can be used to improve asset
allocation and risk management. Therefore, volatility spillover, co-movement
and contagion of financial markets have been extensively investigated in the
literature.

Researchers typically adopted a vector autoregressive model, a multivariate
generalized autoregressive conditional heteroskedasticity (GARCH) model or a
combination of both models to analyze volatility spillover, co-movement and
contagion of financial markets (Baele (2005), Dungey et al. (2005), Forbes
and Rigobon (2002),\ Karolyi (1995), King et al (1994) and the references
therein). Additionally, a copula model or a combination of a copula and an
existing multivariate model has been used to investigate dependence between
financial markets (Garcia and Tsafack (2011), Lee and Long (2009), and
Rodriguez (2007), among others).

While these existing methods generally depend on parametric modeling of
conditional variance, conditional correlation or copula of multivariate
financial time series, recently researchers introduced some methods that do
not require any modeling and focus directly on the quantile dependence of
financial time series (Barun\'{\i}k and Kley (2015), Cappiello et al (2014),
Han et al (2016), Li et al (2015) and Schmitt et al (2015), among others).
These works provide various new methods to measure quantile dependence that
is not captured by classical measures based on linear correlation. Some
methods such as that used in Cappiello et al (2014) test contagion or
constant correlation between financial time series, which can provide useful
implications for asset allocation. However, little research has explored
beyond basic measurement of quantile dependence between financial time
series to investigate how to directly make use of measured quantile
dependence in volatility forecasting, asset allocation and/or risk
management.

The main motivation of this paper is to address this gap. We first measure
detailed quantile dependence between stock markets and examine
quantile-based directional predictability between stock markets. Using the
quantile-based dependence and directional predictability, we introduce and
evaluate a method to improve volatility forecasting in each stock market.

We consider the daily S\&P 500 index, FTSE 100 index, DAX index, CAC 40
index and Nikkei 250 index and examine quantile dependence between the US
stock return and stock return series for the UK, Germany, France and Japan,
i.e. quantile dependence between US-UK, US-Germany, US-France or US-Japan
bivariate stock market returns. To examine detailed quantile-based
relationships between stock markets, we adopt the cross-quantilogram
recently proposed by Han et al (2016). The cross-quantilogram is a
correlation statistic of quantile hit processes and measures dependence
between the quantile range of one time series and the quantile range of the
other time series. Therefore, it can provide quantile-based dependence
between two financial markets. One can set up a cross-quantilogram for
specific quantile ranges of interest or for an arbitrary large lag, and it
is simple to interpret these results.

The results based on the cross-quantilogram show the following. First,
negative spillover (left-tail dependence between stock markets) is stronger
than positive spillover (right-tail dependence between stock markets). The
cross-quantilogram has higher values and remains significant for larger lags
when we consider left-tail dependence between stock markets. Second, there
exists stronger quantile dependence or directional predictability from the
US stock market to the UK, Germany, France and Japan markets than the other
way around. Third, when stock returns are devolatized and standardized
residuals are used, directional predictability remains significant only at
the first lag in the tail parts from the US\ market to other markets (UK,
Germany, France or Japan), but it disappears from other markets (UK,
Germany, France or Japan) to the US market.

Using these findings, we consider a simple way to improve volatility
forecasting. In particular, we modify a volatility model to exploit the
quantile-based directional predictability from the US\ market to markets in
the UK, Germany, France and Japan. In a volatility model for stock markets
in the UK, Germany, France and Japan, we introduce an additional
multiplicative component that can be predicted from a tail event in the US\
stock market. We show that such quantile-augmented volatility models provide
superior in-sample and out-of-sample volatility forecasts. We also find that
our multiplicative model provides better volatility forecasts than the usual
additive GARCH-X model.

The rest of the paper is organized as follows. Section 2 explains the
cross-quantilogram and related Box-Ljung type test statistic. Sections 3
provides the data description and results on quantile dependence between
stock markets. It presents results of auto-quantilogram and
cross-quantilogram for stock return series and the standardized residual.
Section 4 presents the application of quantile dependence to volatility
forecasting and Section 5 concludes the paper.

\section{Econometric Tool}

Linton and Whang (2007) introduced the (auto-) quantilogram to measure
dependence in different parts of the distribution of a stationary time
series based on the correlogram of quantile hits. Han et al (2016) developed
a multivariate version called the cross-quantilogram. The cross-quantilogram
can be used 1) to measure quantile dependence between two series, 2) to test
directional predictability between two series, and 3) to test model
specification. They proposed and investigated the stationary bootstrap
procedure and a self-normalized approach to construct the confidence
intervals of the cross-qauntilogram.

As explained in Linton and Whang (2007) and Han et al (2016), the advantages
of the cross-quantilogram are as follows: 1) it is simple to interpret, 2)
no moment condition is required for time series, 3) it captures the
properties of a joint distribution, 4) it can consider arbitrary lags. The
second advantage is particularly important when we use the
cross-quantilogram to analyze financial time series. It is well known that
finite fourth moments do not exist for most stock return or exchange rate
return series due to heavy tails. While commonly used models such as
multivariate GARCH models in general assume the existence of finite fourth
moments of time series, no moment condition is required for the
cross-quantilogram.

We let $q_{i,t}$($\tau _{i}$) be either $\tau _{i}$ conditional or
unconditional quantile of $y_{i,t}.$ The cross-quantilogram measures
dependence between two events $\{y_{1,t}<q_{1,t}(\tau _{1})\}$ and $%
\{y_{2,t-k}<q_{2,t-k}(\tau _{2})\}$ for an arbitrary pair of $\tau =(\tau
_{1},\tau _{2})^{\prime }$ and a positive integer $k$. In the literature, $%
\{1[y_{i,t}<q_{i,t}(\cdot )]\}$ is called the quantile-hit or
quantile-exceedance process for $i=1,2$, where $1[\cdot ]$ denotes the
indicator function.

The cross-quantilogram is the cross-correlation of the quantile-hit
processes and is defined as 
\begin{equation}
\rho _{\tau }\left( k\right) =\frac{E\left[ \psi _{\tau _{1}}\left(
y_{1,t}-q_{1,t}(\tau _{1})\right) \psi _{\tau _{2}}\left(
y_{2,t-k}-q_{2,t-k}(\tau _{2})\right) \right] }{\sqrt{E\left[ \psi _{\tau
_{1}}^{2}\left( y_{1,t}-q_{1,t}(\tau _{1})\right) \right] }\sqrt{E\left[
\psi _{\tau _{2}}^{2}(y_{2,t-k}-q_{2,t-k}(\tau _{2}))\right] }}  \label{q1}
\end{equation}%
for $k=0,\pm 1,\pm 2,\dots ,$ where 
\begin{equation*}
\psi _{\tau _{i}}(y_{i,t}-q_{i,t}(\tau _{i}))=1[y_{i,t}<q_{i,t}(\tau
_{i})]-\tau _{i}.
\end{equation*}%
Its sample counterpart is 
\begin{equation*}
\hat{\rho}_{\tau }(k)=\frac{\sum_{t=k+1}^{T}\psi _{\tau _{1}}(y_{1,t}-\hat{q}%
_{1,t}(\tau _{1}))\psi _{\tau _{2}}(y_{2,t-k}-\hat{q}_{2,t-k}(\tau _{2}))}{%
\sqrt{\sum_{t=k+1}^{T}\psi _{\tau _{1}}^{2}(y_{1,t}-\hat{q}_{1,t}(\tau _{1}))%
}\sqrt{\sum_{t=k+1}^{T}\psi _{\tau _{2}}^{2}(y_{2,t-k}-\hat{q}_{2,t-k}(\tau
_{2}))}},
\end{equation*}%
where $\hat{q}_{i,t}$($\tau _{i}$) is the estimate of either $\tau _{i}$
conditional or unconditional quantile of $y_{i,t}.$ As an example, Figure 1
provides a pair of events: $\{y_{1,t}<q_{1,t}(\tau _{1})\}$ for $\tau
_{1}=0.05$ and $\{y_{2,t-k}<q_{2,t-k}(\tau _{2})\}$ for $\tau _{2}=0.5.$
Given $y_{2,t-k}$ is located below its median, the cross-quantilogram $\rho
_{\tau }(k)$ is zero if the probability of $y_{1,t}$ being located below its 
$0.05$ quantile is the same as $0.05.$

Instead of two events $\{y_{1,t}<q_{1,t}(\tau _{1})\}$ and $%
\{y_{2,t-k}<q_{2,t-k}(\tau _{2})\},$ one may be interested in measuring the
dependence between two events $\{q_{1,t}(\tau _{1}^{l})<y_{1,t}<q_{1,t}(\tau
_{1}^{h})\}$ and $\{q_{2,t-k}(\tau _{2}^{l})<y_{2,t-k}<q_{2,t-k}(\tau
_{2}^{h})\}$ for arbitrary quantile ranges $\left[ \tau _{1}^{l},\tau
_{1}^{h}\right] $ and $\left[ \tau _{2}^{l},\tau _{2}^{h}\right] $. Figure
2\ provides various events $\{q_{i,t}(\tau _{i}^{l})<y_{i,t}<q_{i,t}(\tau
_{i}^{h})\}$ for different quantiles for $\tau _{i}^{l}$ and $\tau _{i}^{h}.$
To obtain the dependence of such events, one can use an alternative version
of the cross-quantilogram that is defined by replacing $\psi _{\tau
_{i}}(y_{it}-q_{i,t}(\tau _{i}))$ in (\ref{q1}) with%
\begin{equation*}
\psi _{\left[ \tau _{i}^{l},\tau _{i}^{h}\right] }(y_{it}-q_{i,t}(\left[
\tau _{i}^{l},\tau _{i}^{h}\right] ))=1[q_{i,t}(\tau
_{i}^{l})<y_{it}<q_{i,t}\left( \tau _{i}^{h}\right) ]-\left( \tau
_{i}^{h}-\tau _{i}^{l}\right) .
\end{equation*}%
See footnote 4 in Han et al (2016). This alternative version could be easier
to interpret and therefore we will adopt this alternative version of the
cross-quantilogram in this paper.

If $\rho _{\tau }(k)=0,$ there is no dependence or directional
predictability from an event $\{q_{2,t-k}(\tau _{2}^{l})\leq y_{2,t-k}\leq
q_{2,t-k}(\tau _{2}^{h})\}$ to an event $\{q_{1,t}(\tau _{1}^{l})\leq
y_{1,t}\leq q_{1,t}(\tau _{1}^{h})\}.$ If $\rho _{\tau }(k)\neq 0,$ there
exists quantile dependence or directional predictability between two events.
If $\rho _{\tau }(k)>0,$ it is more likely for $y_{1,t}$ to be located in
the range $[q_{1,t}(\tau _{1}^{l}),q_{1,t}(\tau _{1}^{h})]$ when $y_{2,t-k}$
is located in the range $[q_{2,t-k}(\tau _{2}^{l}),q_{2,t-k}(\tau
_{2}^{h})]. $ If $\rho _{\tau }(k)<0,$ it is less likely for $y_{1,t}$ to be
located in the range $[q_{1,t}(\tau _{1}^{l}),q_{1,t}(\tau _{1}^{h})]$ when $%
y_{2,t-k}$ is located in the range $[q_{2,t-k}(\tau _{2}^{l}),q_{2,t-k}(\tau
_{2}^{h})]. $ The stationary bootstrap inference procedure is still valid
for this alternative version, as mentioned in Han et al (2016) and,
therefore, we will use it to construct confidence bands.

Using the cross-quantilogram, we can conduct related Portmanteau tests.
Suppose that $\tau \in \mathcal{T}$ and $p$ are given. One may be interested
in testing 
\begin{eqnarray*}
H_{0} &:&\rho _{\tau }(1)=\dots =\rho _{\tau }(p)=0, \\
H_{1} &:&\rho _{\tau }(k)\not=0\text{ for some }k\in \{1,\dots ,p\}.
\end{eqnarray*}%
For this test, the Box-Pierce type test statistic $\hat{Q}_{\tau
}^{(p)}=T\sum_{k=1}^{p}\hat{\rho}_{\tau }^{2}(k)$ can be used. We will use
the Box-Ljung version $\check{Q}_{\tau }^{(p)}=T\left( T+2\right)
\sum_{k=1}^{p}\hat{\rho}_{\tau }^{2}\left( k\right) \left/ \left( T-k\right)
\right. $ in this paper because it has better finite sample performance for
a large $p$ and a small sample size. Han et al (2016) also analyze the
sup-version test statistic over a set of quantiles and the partial
cross-quantilogram.

\section{Quantilogram Analysis}

\subsection{Data and Setup}

We investigate quantile dependence and directional predictability between
the US stock market and stock markets in the UK, Germany, France and Japan,
i.e. quantile dependence and directional predictability between US-UK,
US-Germany, US-France and US-Japan bivariate stock market returns. We
consider the daily S\&P 500 index, FTSE 100 index, DAX index, CAC 40 index
and Nikkei 250 index. To calculate the cross-quantilogram between the US
stock return and the stock return series for the UK, Germany, France and
Japan, we only consider days $t$ for which we have observations from both
indices for each pair. The sample period and sample size for each pair of
indices is given in Table 1.\footnote{%
The data set is from realized library 0.1 by the Oxford-Man Institute.} We
consider samples until the end of 2007 so that strict stationarity holds for
the data. We demean each stock return series by subtracting its sample mean.

We let $\tau _{i}$ denote a quantile range $\left[ \tau _{1}^{l},\tau
_{1}^{h}\right] $ in this section. The quantile range of stock return $\tau
_{i}$ is set to be [0,0.05], [0.05,0.1], [0.1,0.2], [0.2,0.4], [0.4,0.6],
[0.6,0.8], [0.8,0.9], [0.9,0.95] or [0.95,1]. We first let $\tau _{1}=\tau
_{2}$ for the next two subsections and consider the case with $\tau _{1}\neq
\tau _{2}$ later. We let lag $k=1,\dots ,20.$ We use the stationary
bootstrapping procedure by Politis and Romano (1994) to obtain confidence
intervals based on 1,000 bootstrap replicates. The tuning parameter is
chosen by adapting the rule suggested by Politis and White (2004) (and later
corrected in Patton et al. (2009)).

\subsection{Auto-Quantilogram and Cross-Quantilogram}

We first examine the auto-quantilogram in the US stock market and the UK
stock market. The results for the German, French or Japanese stock market
are in general similar to those for the UK stock market and, therefore, we
do not include them in the paper. Figures 3(a) and 3(b) show the
auto-quantilogram and the Box-Ljung test statistic for the S\&P 500 index
return series. The auto-quantilogram is significantly positive at some lags
for $\tau _{1}$=[0,0.05], [0.4,0.6] or [0.95,1.0], which makes the Box-Ljung
test statistic in Figure 3(b) significant for the same quantile range $\tau
_{1}.$

Figures 4(a) and 4(b) present the auto-quantilogram and the Box-Ljung test
statistic for the FTSE 100 index return series. The results of the UK stock
market are in general similar to those of the US stock market. For both tail
parts ($\tau _{1}$=[0,0.05] or [0.95,1.0]) and the mid-range ($\tau _{1}$%
=[0.4,0.6]), the auto-quantilogram is significantly positive for some lags.

Next, we investigate the cross-quantilogram between the US stock market and
the UK stock market. Figures 5(a) and 5(b) provide the cross-quantilogram
and the Box-Ljung test statistic from the US stock market to the UK stock
market, i.e., $y_{1,t}$ is the FTSE 100 index return and $y_{2,t-k}$ is the
S\&P 500 index return. This shows that there exists directional
predictability from the US market to the UK\ market for various quantile
ranges. When we consider only the first lag, $k=1$, the cross-quantilogram
is significantly positive for $\tau _{1}$=[0,0.05], [0.05,0.1], [0.1,0.2],
[0.9,0.95] or [0.95,1.0].

It is not surprising to note that the quantile dependence is asymmetric. For
the left-tail ($\tau _{1}$=[0,0.05]), the cross-quantilogram exhibits much
higher values and it is significant for larger lags. This implies that when
there is a very large negative loss in the US stock market, it is more
likely that there is also a very large loss in the UK stock market for quite
a long time. Table 3 provides the value of $\hat{\rho}_{\tau _{1}}(1),$ the
cross-quantilogram at the first lag, for both tail parts; it is 0.25 for the
left-tail ($\tau _{1}$=[0,0.05]) and 0.13 for the right-tail ($\tau _{1}$%
=[0.95,1.0]). This implies that the negative spillover (risk spillover) is
stronger than the positive spillover. Such an asymmetric relationship is in
accordance with what we commonly observe in international stock markets.

Figures 6(a) and 6(b) present the cross-quantilogram and the Box-Ljung test
statistic from the UK stock market to the US stock market, i.e., $y_{1,t}$
is the S\&P 500 index return and $y_{2,t-k}$ is the FTSE 100 index return.
Compared to the results in Figures 5(a) and 5(b), the dependence is much
weaker. The cross-quantilogram in general has a lower value and is
significant at some lags only for $\tau _{1}$=[0,0.05], [0.4,0.6] or
[0.95,1.0]. The cross-quantilogram from the UK market to the US market
exhibits similar patterns to the auto-quantilogram for the US market in
Figure 3(a).

\subsection{Results of Devolatized Return Series}

The results in the previous subsection show that dependence or
predictability still exists from the UK stock market to the US stock market
despite it being much weaker than the case from the US market to the UK
market. However, the auto-quantilogram in the US market exhibits similar
patterns to the cross-quantilogram from the UK market to the US market,
while it is obviously different from the cross-quantilogram from the US
market to the UK market. Therefore, the quantile dependence from the UK
stock return to the US stock return could be an artifact due to persistence
and synchronicity in the marginal volatilities of the two stock return
series. As discussed in Section 3 in Davis et al (2013), this phenomenon is
similar to the well-known issue with the cross-correlation function of
linear bivariate time series. Unless one or all time series are whitened,
the cross-correlation may appear to be spuriously significant (see Chapter
11 in Brockwell and Davis (1991)).

Hence, in this subsection, we devolatilize each stock return series and
examine the cross-quantilogram using standardized residuals. For each return
series, we estimate the GJR-GARCH(1,1) model:%
\begin{eqnarray*}
y_{i,t} &=&\sigma _{t}\varepsilon _{t}, \\
\sigma _{t}^{2} &=&\omega +\alpha y_{i,t-1}^{2}+\gamma
y_{i,t-1}^{2}I(y_{i,t-1}<0)+\beta \sigma _{t-1}^{2}.
\end{eqnarray*}%
We adopt the GJR-GARCH model to accommodate the asymmetric relationship
between stock return and volatility. The innovation $\varepsilon _{t}$ is is
assumed to be iid (0,1) and therefore the standardized residual $\hat{%
\varepsilon}_{t}=y_{i,t}/\hat{\sigma}_{t}$ is presumed to be serially
uncorrelated. Testing serial correlation in the standardized residual is one
of the most common ways to check model specification in the literature.
Table 2 reports the `usual' Ljung-Box Q-statistic based on auto-correlations
of $\hat{\varepsilon}_{t}$ or $\hat{\varepsilon}_{t}^{2}.$ For all stock
return series, the p-values of the Ljung-Box Q-statistic for lag 10 or 20
are larger than 0.05. This shows that $\hat{\varepsilon}_{t}$ and $\hat{%
\varepsilon}_{t}^{2}$ are serially uncorrelated and suggests that the
GJR-GARCH model is an appropriate volatility model for this return series.

Now we use the standardized residual instead of the stock return series and
conduct quantilogram analysis. Figures 7(a)-8(b) provide the
auto-quantilogram and the Box-Ljung test statistic using the standardized
residual $\hat{\varepsilon}_{t}$ for the US market or the UK\ market. The
auto-quantilogram is insignificant in most cases for both stock markets,
which is in accordance with the results of the `usual' Ljung-Box Q-statistic
on $\hat{\varepsilon}_{t}$ and $\hat{\varepsilon}_{t}^{2}$ in Table 2 and
suggests the GJR-GARCH model is appropriate for modeling each stock return
series.

Figures 9(a) and 9(b) present the cross-quantilogram and the Box-Ljung test
statistic from the US market to the UK market using the standardized
residual, i.e. $y_{1,t}$ is $\hat{\varepsilon}_{t}$ for the FTSE 100 index
return and $y_{2,t-k}$ is $\hat{\varepsilon}_{t-k}$ for the S\&P 500 index
return. The cross-quantilogram has a large positive value at the first lag
for the left-tail ($\tau _{1}$=[0,0.05]), while it is mostly insignificant
in the rest of the cases. Even after devolatizing the returns series, there
still exists directional predictability from the US market to the UK market
in the left-tail. Figures 10(a) and 10(b) provide the cross-quantilogram and
the Box-Ljung test statistic from the UK market to the US market using the
standardized residual, i.e. $y_{1,t}$ is $\hat{\varepsilon}_{t}$ for the
S\&P 500 index return and $y_{2,t-k}$ is $\hat{\varepsilon}_{t-k}$ for the
FTSE 100 index return. The cross-quantilogram is insignificant in almost all
cases and consequently the Box-Ljung test statistic is insignificant in all
cases.

When we devolatize only one stock return series, the results are in general
similar. For example, when $y_{1,t}$ is $\hat{\varepsilon}_{t}$ for the FTSE
100 index return and $y_{2,t-k}$ is the S\&P 500 index return itself,
predictability still exists at the first lag from the US market to the UK
market. However, when $y_{1,t}$ is $\hat{\varepsilon}_{t}$ for the S\&P 500
index return and $y_{2,t-k}$ is the FTSE 100 index return itself, no
predictability exists from the UK market to the US market.

When one or both stock return series is devolatized, directional
predictability still appears from the US market to the UK\ market in the
left tail, but disappears from the UK market to the US market in all
quantile ranges. This could be due to the dominance of the US stock market.
Another possibility is the difference in stock market opening times. The
stock market opening times are Japan (00:00-06:00), UK/Germany/France
(08:00-16:30) and US (14:30-21:00) in GMT. There are two hours of overlap
between the European and US stock market opening times. One may surmise that
a shock in the UK market on day $t$ will be transmitted to the US market on
the same day and, consequently, directional predictability will disappear
from the UK market to the US\ market at the first lag. However, this does
not make sense considering that the US-Japan case presented in Tables 2 and
3 shows similar results as the US-UK case despite no overlap between the US
and Japan stock market opening times. We conjecture that the market
dominance of the US causes large significant values of the
cross-quantilogram for the first lag from the US market to each stock market
in Europe and Japan.

When we replace the UK stock market with the German or French stock market,
the cross-quantilogram exhibits similar patterns as the US-UK case. Table 3
provides the cross-quantilograms at the first lag from the US stock market
to each stock market and Table 4 presents those from each stock market to
the US stock market. For example, when we consider the US-Germany case, we
observe the following: 1) dependence is stronger for the case from the US
market to the German market than the other way around, 2) the negative
spillover is stronger than the positive spillover, 3) when the standardized
residual is used, directional predictability still exists in both tails from
the US market to the German market, but disappears from the German market to
the US market.

There is an interesting difference in the US-Japan case. The positive
spillover from the US market to the Japanese market is stronger than the
negative spillover, i.e., $\hat{\rho}_{\tau }(1)=0.18$ for $\tau _{1}$%
=[0,0.05] and $\hat{\rho}_{\tau }(1)=0.21$ for $\tau _{1}$=[0.95,1.0],
whereas the negative spillover is stronger from the US market to\ three
European markets. Figures 11(a) and 11(b) present the cross-quantilogram
from the US\ market to the Japanese market. In Figure 11(a), the
cross-quantilogram is also significantly positive at the first lag for $\tau
_{1}$=[0.9,0.95] while it is insignificant for $\tau _{1}$=[0.05,0.1]. When
we use the standardized residuals from the GJR-GARCH model, Figure 11(b)
shows that the cross-quantilogram is significantly positive for both tails
at the first lag.

\subsection{Results of Cross-Quantile Ranges}

Instead of letting $\tau _{1}=\tau _{2},$ we now consider the case with $%
\tau _{1}\neq \tau _{2}.$ We let the quantile range of the US\ stock market $%
\tau _{2}$ be either [0,0.05] or [0.95,1.0]. We set the quantile range of
the UK stock market $\tau _{1}$ to be [0,0.05], [0.05,0.1], [0.1,0.2],
[0.2,0.4], [0.4,0.6], [0.6,0.8], [0.8,0.9], [0.9,0.95] or [0.95,1] as in
previous subsections.

First, we examine dependence and directional predictability from the
left-tail event in the US market to various quantile ranges of the UK stock
market. Figure 12(a) presents the cross-quantilogram from the US market to
the UK\ market, i.e., $y_{1,t}$ is the FTSE 100 index return, $y_{2,t-k}$ is
the S\&P 500 index return and $\tau _{2}$=[0,0.05]. The first plot in the
first row in Figure 12(a) is identical to that in Figure 5(a) where $\tau
_{1}=\tau _{2}=$[0,0.05]. For mid-quantile ranges of the UK market ($\tau
_{1}$=[0.2,0.4], [0.4,0.6] or [0.6,0.8]), the cross-quantilogram is
significantly negative for some lags. This means that it is less likely for
the UK stock return to be located in mid-quantile ranges when there is a
large loss in the US market at day $t-k.$ For the right-tail of the UK stock
market ($\tau _{1}$=[0.95,1]), the cross-quantilogram is close to zero and
insignificant at the first lag but it is mostly significantly positive from
the second lag to the last lag. This could be due to the bouncing effect
after a large negative shock. It is interesting to note that values of the
cross-quantilogram are higher in the right-tail than in the left-tail from
the second lag, while the value is very high in the left-tail only at the
first lag.

Second, we consider the case from the right-tail event in the US market.
Figure 12(b) presents the cross-quantilogram from the US market to the UK\
market, i.e., $y_{1,t}$ is the FTSE 100 index return, $y_{2,t-k}$ is the
S\&P 500 index return and $\tau _{2}$=[0.95,1]. The last plot in the third
row in Figure 12(b) is identical to that in Figure 5(a) where $\tau
_{1}=\tau _{2}=$[0.95,1]. In general, the dependence is weaker than the case
in Figure 12(a). On various quantile ranges of the UK stock market return, a
large negative shock in the US stock market has a stronger influence than a
large positive shock. For $\tau _{1}$=[0.9,0.95], the cross-quantilogram is
significantly positive at the first lag. The figure shows that, when there
is a large gain in the US\ stock market, it is more likely for the UK\ stock
market to have a large or a relatively large gain on the next day.

Next, we use the standardized residuals from the GJR-GARCH model and examine
the same cross-quantile range aspects. When the standardized residuals are
used, the cross-quantilogram is mostly insignificant except for some
quantile ranges at the first lag. Figure 13(a) considers the left-tail case
corresponding to Figure 12(a). At the first lag, the cross-quantilogram is
significantly positive for $\tau _{1}$=[0,0.05], [0.05,0.1] or [0.1,0.2].
Figure 13(b) presents the right-tail case corresponding to Figure 12(b). The
cross-quantilogram is mostly close to zero and insignificant.

\section{Application in Volatility Forecasting}

\subsection{Quantile-Augmented Volatility Model}

In this section, we consider a method that uses the findings in the previous
section to improve volatility forecasting. There exists directional
predictability from a tail event in $y_{2,t-1}$, i.e., US stock return at
day $t-1$ for $\tau _{2}=[0,0.05]$ and $[0.95,1]$, to the standardized
residual $\hat{\varepsilon}_{t}$ in markets in the UK, Germany, France and
Japan. This result suggests that we can decompose $\hat{\varepsilon}_{t}$
into two parts such that $\hat{\varepsilon}_{t}=\sqrt{\hat{f}_{t}}\hat{\eta}%
_{t}$ where $f_{t}$ is the predictable component from a tail event in $%
y_{2,t-1}$ and $\eta _{t}$ is an unpredictable component. Using this idea,
we consider the following volatility model for stock return series $y_{1,t}$
of each market in the UK, Germany, France and Japan;%
\begin{equation*}
y_{1,t}=\sqrt{h_{t}f_{t}}\eta _{t}
\end{equation*}%
where $h_{t}$ is the GJR-GARCH model, $f_{t}$ is a function of a tail event
in $y_{2,t-1}$ and $\eta _{t}$ is $iid$ (0,1). The return series in each
market has three multiplicative components. The first component $h_{t}$ is a
function of past values of $y_{1,t}$ and it is possible to specify it as
anther GARCH-type model. We call $h_{t}$ a base volatility model. We can
model the second component $f_{t}$ in various ways including nonparametric
methods. In this paper, we consider the following simple specification: 
\begin{equation}
f_{t}\left( \delta \right) =\delta _{0}+\delta
_{1}y_{2,t-1}^{2}I(y_{2,t-1}\leq q_{2}\left( 0.05\right) )+\delta
_{2}y_{2,t-1}^{2}I(y_{2,t-1}\geq q_{2}\left( 0.95\right) )  \label{ft}
\end{equation}%
where $y_{2,t}$ is the return series in the US stock market and $q_{2}\left(
0.05\right) $ or $q_{2}\left( 0.95\right) $ is 0.05 or 0.95 quantile of $%
y_{2,t}$, respectively.

In this manner, the conditional variance of $y_{1,t}$ is augmented as 
\begin{equation*}
\sigma _{t}^{2}=h_{t}\times f_{t}
\end{equation*}%
where $h_{t}$ is a base volatility model such as the GJR-GARCH model and $%
f_{t}$ is defined as in (\ref{ft}). If the stock index return in US is below
the 5\% quantile or above the 95\% quantile, a positive value of $\delta
_{1} $ or $\delta _{2}$ will make the volatility in each stock market higher
on the next day. We call this model the \textit{quantile-augmented}
volatility model (QA model).

Another way to accommodate the directional predictability from the US market
to each stock market in volatility modeling is to adopt the following
additive GARCH-X model; 
\begin{equation}
y_{1,t}=\sqrt{h_{t}}\eta _{t}  \label{garchx1}
\end{equation}%
where 
\begin{multline*}
h_{t}=\omega +\alpha y_{1,t-1}^{2}+\gamma y_{1,t-1}^{2}I(y_{t-1}<0)+\beta
h_{t-1} \\
+\delta _{1}y_{2,t-1}^{2}I(y_{2,t-1}\leq q_{2}\left( 0.05\right) )+\delta
_{2}y_{2,t-1}^{2}I(y_{2,t-1}\geq q_{2}\left( 0.95\right) )
\end{multline*}%
and $\eta _{t}$ is $iid$ (0,1). The GARCH-X model is a typical way to
accommodate exogenous covariates in volatility modeling (see Han and
Kristensen (2015) and references therein).

Now we discuss the estimation method of the QA model. We can rearrange the
model 
\begin{equation*}
y_{1,t}=\sqrt{h_{t}\left( \theta \right) f_{t}\left( \delta \right) }\eta
_{t}\text{ for }\eta _{t}\sim iid(0,1),
\end{equation*}%
into%
\begin{equation*}
y_{t}^{2}/h_{t}\left( \theta \right) =f_{t}\left( \delta \right) +u_{t}
\end{equation*}%
where $u_{t}=f_{t}\left( \delta \right) \left( \eta _{t}^{2}-1\right) .$
Here $u_{t}$ is a Martingale difference sequence$.$ The estimation procedure
is as follows:

\begin{enumerate}
\item Estimate $\theta $ in the base model $h_{t}\left( \theta \right) ,$
for example the GJR-GARCH model, using the quasi-maximum likelihood
estimation (QMLE) method from%
\begin{equation*}
y_{1,t}=\sqrt{h_{t}\left( \theta \right) }\varepsilon _{t}\text{ for }%
\varepsilon _{t}\sim iid(0,1).
\end{equation*}

\item Rescale the squared return and estimate $\delta $ in the following
model using the OLS method%
\begin{equation*}
y_{t}^{2}/h_{t}(\hat{\theta})=f_{t}\left( \delta \right) +u_{t}.
\end{equation*}

\item Using the estimates from the previous steps and obtain%
\begin{equation*}
\hat{\sigma}_{t}^{2}=h_{t}(\hat{\theta})\times f_{t}(\hat{\delta}).
\end{equation*}
\end{enumerate}

\subsection{Forecast Evaluation Method}

We evaluate the within-sample and out-of-sample predictive power of the
quantile-augmented volatility model. We will compare the within-sample and
out-of-sample forecasts of the base model (GJR-GARCH, $\hat{\sigma}%
_{t,base}^{2}=\hat{h}_{t}$) and the QA model ($\hat{\sigma}_{t,QA}^{2}=\hat{h%
}_{t}\times \hat{f}_{t}$). To evaluate the volatility forecast, we adopt the
following standard procedure. First, we use the realized kernel as a proxy
for actual volatility. Barndorff-Nielsen et al. (2008) introduced the
realized kernel and it has some robustness to market microstructure noise.
The realized kernels of the return series are available in the `Oxford-Man
Institute's realised library' database produced by Heber \textit{et al.}
(2009).

Second, we use the QLIKE loss function defined as%
\begin{equation}
L(\hat{\sigma}_{t}^{2},\sigma _{t}^{2})=\frac{\sigma _{t}^{2}}{\hat{\sigma}%
_{t}^{2}}-\log \frac{\sigma _{t}^{2}}{\hat{\sigma}_{t}^{2}}-1  \label{qlike}
\end{equation}%
where $\left( \sigma _{t}^{2}\right) $ is the proxy for actual volatility
and $\left( \hat{\sigma}_{t}^{2}\right) $ is the within-sample or
out-of-sample volatility forecast. Even if realized measures are known to be
better measures, they are imperfect and noisy proxies for actual volatility.
There has been research on loss functions that are robust to the use of a
noisy volatility proxy (see Hansen and Lunde (2006), Patton (2010) and
Patton and Sheppard (2009)). Patton (2010) shows that the QLIKE loss
function is robust and, in particular, Patton and Sheppard (2009) show in
their simulation study that the QLIKE loss function has the highest power.

Third, the significance of any difference in the QLIKE\ loss is tested via a
Diebold-Marinao and West (DMW) test. See Diebold-Marinao (1995) and West
(1996). A DMW statistic is computed using the difference in the losses of
two models 
\begin{eqnarray}
d_{t} &=&L(\hat{\sigma}_{t,base}^{2},\sigma _{t}^{2})-L(\hat{\sigma}%
_{t,QA}^{2},\sigma _{t}^{2})  \notag \\
DMW_{T} &=&\frac{\sqrt{T}\bar{d}_{T}}{\sqrt{\widehat{avar}\left( \sqrt{T}%
\bar{d}_{T}\right) }}  \label{DMW}
\end{eqnarray}%
where $\bar{d}_{T}$ is the sample mean of $d_{t}$ and $T$ is the number of
forecasts. The asymptotic variance of the average is computed using a
Newey-West variance estimator with the number of lags set to $\left[ T^{1/3}%
\right] .$ If $DMW_{T}$ is positive, it means that our quantile-augmented
model has a smaller loss than the base model. The DMW test for equal
predictability is for%
\begin{equation*}
H_{0}:\mathbb{E}\left[ d_{t}\right] =0
\end{equation*}%
and the asymptotic distribution of the test statistic is standard normal
under the null hypothesis.

\subsection{Forecast Evaluation Results}

We first compare fitted values of volatility for the entire sample period.
Table 5 shows the DMW test results for each series. In all cases, the DMW
test statistics are positive and statistically significant at the 1\% level.
This shows that our quantile-augmented model significantly outperforms the
GJR-GARCH model.

Next we compare one-step ahead out-of-sample forecasts. We adopt the rolling
window procedure with a moving window of eight years (2016 days) and produce
one-step ahead out-of-sample forecasts. The forecast period and number of
forecasts for each series are given in Table 6.

Table 5 shows the DMW test results for the out-of-sample forecasts. The
results are similar to those for the in-sample comparison. The
quantile-augmented model significantly outperforms the GJR-GARCH model. Both
in-sample and out-of-sample comparison results show that a simple augmented
model using quantile dependence and directional predictability from the US
market can significantly improve volatility forecasting.\bigskip

\textbf{Remark 1}: Instead of a tail event in $y_{2,t-1},$ one may use a
tail event in $\hat{\varepsilon}_{2,t-1}$ that is the standardized residual
of the GJR-GARCH model for $y_{2,t-1}.$ Accordingly, we can adjust $f_{t}$
as follows:%
\begin{equation*}
f_{t}\left( \delta \right) =\delta _{0}+\delta _{1}\hat{\varepsilon}%
_{2,t-1}^{2}I(\hat{\varepsilon}_{2,t-1}\leq q_{2}\left( 0.05\right) )+\delta
_{2}\hat{\varepsilon}_{2,t-1}^{2}I(\hat{\varepsilon}_{2,t-1}\geq q_{2}\left(
0.95\right) )
\end{equation*}%
where $q_{2}\left( 0.05\right) $ or $q_{2}\left( 0.95\right) $ are 0.05 or
0.95 quantile of $\hat{\varepsilon}_{2,t}$, respectively. We still obtain
similar results. For all cases, the quantile-augmented model significantly
outperforms the base model in both in-sample and out-of-sample forecasts.

\textbf{Remark 2}: We consider two different base models instead of the
GJR-GARCH model and conduct the same in-sample and out-of-sample forecast
evaluations. One is the GJR-GARCH model with $t$-distribution, in which the
innovation $\varepsilon _{t}$ follows the $t$-distribution. The other is the
HEAVY model by Shephard and Sheppard (2010). Specifically, we use their
HEAVY-r model:%
\begin{eqnarray*}
y_{1,t} &=&\sigma _{t}\varepsilon _{t} \\
\sigma _{t}^{2} &=&\omega +\beta \sigma _{t-1}^{2}+\pi RM_{1,t-1}
\end{eqnarray*}%
where $RM_{1,t}$ is the realized volatility measure of $y_{1,t}$ at time $t.$
Shephard and Sheppard (2010) and Hansen et al (2012) show that this GARCH-X
type model using a realized volatility measure as the covariate performs
better than the standard GARCH model. Following Shephard and Sheppard
(2010), we use the realized kernel as $RM_{1,t}$. Tables 7 and 8 show the
results of in-sample and out-of-sample forecast comparisons using the
alternative base models. They show that the quantile-augmented approach
still significantly improves volatility forecasting.

\textbf{Remark 3}: We consider the additive GARCH-X model given in (\ref%
{garchx1}). When we compare in-sample forecasts, the additive GARCH-X model
provides lower QLIKE\ losses than the GJR-GARCH model, but the DMW test
statistics are in general insignificant. This shows that the additive
GARCH-X model is not as effective as the multiplicative approach in our
model.

\textbf{Remark 4}: We apply the same quantile-augmented approach in
volatility modeling of the US\ stock return. For the US stock return $%
y_{2,t} $, we consider%
\begin{equation*}
y_{2,t}=\sqrt{h_{t}f_{t}}\eta _{t}
\end{equation*}%
where $h_{t}$ is the GJR-GARCH model, $\eta _{t}$ is $iid$ (0,1) and 
\begin{equation*}
f_{t}=\delta _{0}+\delta _{1}y_{1,t-1}^{2}I(y_{1,t-1}\leq q_{1}\left(
0.05\right) )+\delta _{2}y_{1,t-1}^{2}I(y_{1,t-1}\geq q_{1}\left(
0.95\right) ).
\end{equation*}%
$y_{1,t}$ is the stock return of one of the markets in the UK, Germany,
France and Japan, and $q_{1}\left( 0.05\right) $ and $q_{1}\left(
0.95\right) $ are the 0.05 and 0.95 quantile of $y_{1,t},$ respectively.
Since the cross-quantilogram analysis in Section 3 shows that there is no
quantile dependence or directional predictability from each market (UK,
Germany, France or Japan) to the US market after devolatizing, there is no
reason to expect that the quantile-augmented model outperforms the base
model in this case. When we compare in-sample forecasts, the
quantile-augmented model does not provide any significant improvement: DMW
test statistics are either insignificantly positive or significantly
negative. This confirms that the quantile-augmented approach should be based
on the quantile dependence or directional predictability revealed in
cross-quantilogram analysis.

\section{Conclusion}

The paper examines quantile dependence and directional predictability
between international stock markets and investigates how to apply these
measures in volatility forecasting. We consider dependence between the US
stock return and stock return series in the UK, Germany, France and Japan,
i.e., quantile dependence between US-UK, US-Germany, US-France and US-Japan
bivariate stock market returns. The results based on the cross-quantilogram
show that the negative spillover is in general much stronger than the
positive spillover, while, exceptionally, positive spillover is stronger in
the case from the US\ stock market to the Japanese stock market. We apply
the cross-quantilogram on standardized residuals as well as stock return
series. There exists directional predictability from the US stock\ market to
markets in the UK, Germany, France and Japan. In particular, tail events in
the US\ stock market influence these stock markets. However, when
standardized residuals are used, there is no directional predictability from
markets in the UK, Germany, France and Japan to the US\ market. Using these
results on quantile dependence and directional predictability, we consider a
simple method to improve volatility forecasting in stock markets in the UK,
Germany, France and Japan. The quantile-augmented volatility model
significantly improves both in-sample and out-of-sample volatility
forecasting, which is robust to the choice of a base volatility model.

Recently, researchers have developed various methods to measure quantile
dependence between time series. This paper considers a simple method to make
use of quantile dependence in order to improve volatility forecasting. The
information provided on detailed quantile dependence can be used for various
purposes, such as modelling univariate or multivariate volatility and
estimating value at risk. More sophisticated methods will be needed to
exploit quantile dependence in asset allocation and risk management.

\pagebreak

\appendix{}

\section{Tables and Figures}

\singlespacing

{\small \noindent }Table 1. Sample period and sample size for each pair of
stock return series\bigskip

\begin{center}
\begin{tabular}{l|l}
\hline
Pair of indices & Sample period (sample size) \\ \hline
FTSE - S\&P 500 & 21 Oct. 1997 - 31 Dec. 2007 (2470) \\ 
DAX - S\&P 500 & 3 Jan. 1996 - 28 Dec. 2007 (2907) \\ 
CAC - S\&P 500 & 3 Jan. 1996 - 31 Dec. 2007 (2908) \\ 
Nikkei - S\&P 500 & 8 Jan. 1996 - 27 Dec. 2007 (2763) \\ \hline
\end{tabular}
\end{center}

\bigskip

\bigskip

{\small \noindent }Table 2. Results of the `usual' Ljung-Box
Q-statistic\bigskip

\begin{center}
\begin{tabular}{llllllllllll}
\hline
&  &  & S\&P &  & FTSE &  & DAX &  & CAC &  & Nikkei \\ \hline
$\hat{\varepsilon}_{t}$ & p-value of Q(10) &  & 0.19 &  & 0.67 &  & 0.99 & 
& 0.11 &  & 0.63 \\ 
& p-value of Q(20) &  & 0.17 &  & 0.55 &  & 0.86 &  & 0.39 &  & 0.80 \\ 
\hline
$\hat{\varepsilon}_{t}^{2}$ & p-value of Q(10) &  & 0.81 &  & 0.59 &  & 0.08
&  & 0.31 &  & 0.39 \\ 
& p-value of Q(20) &  & 0.70 &  & 0.28 &  & 0.19 &  & 0.25 &  & 0.35 \\ 
\hline
\end{tabular}
\end{center}

{\small \noindent Note: The table reports the Ljung-Box Q-statistic on }$%
\hat{\varepsilon}_{t}${\small \ or }$\hat{\varepsilon}_{t}^{2},${\small \
where }$\hat{\varepsilon}_{t}${\small \ is the standardized resisidual from
the GJR-GARCH model.}

\bigskip

\bigskip

{\small \noindent }Table 3. Cross-quantilograms at the first lag from the US
market to other markets\bigskip

\begin{center}
\begin{tabular}{llllllllll}
\hline
& $\tau _{1}$ $(=\tau _{2})$ &  & FTSE &  & DAX &  & CAC &  & Nikkei \\ 
\hline
Return & $[0,0.05]$ &  & 0.25$^{\ast }$ &  & 0.17$^{\ast }$ &  & 0.20$^{\ast
}$ &  & 0.18$^{\ast }$ \\ 
& $[0.95,1]$ &  & 0.13$^{\ast }$ &  & 0.12$^{\ast }$ &  & 0.11$^{\ast }$ & 
& 0.21$^{\ast }$ \\ 
&  &  &  &  &  &  &  &  &  \\ 
Std. residual & $[0,0.05]$ &  & 0.16$^{\ast }$ &  & 0.14$^{\ast }$ &  & 0.14$%
^{\ast }$ &  & 0.14$^{\ast }$ \\ 
& $[0.95,1]$ &  & 0.04 &  & 0.06$^{\ast }$ &  & 0.03 &  & 0.15$^{\ast }$ \\ 
\hline
\end{tabular}
\end{center}

{\small \noindent Note: The table reports }${\small \hat{\rho}}_{\tau }%
{\small (1),}${\small \ a sample cross-quantilogram at the first lag, from
the US stock market to other stock markets, i.e., }$y_{1,t}${\small \ is the
return series of FTSE, DAX, CAC or Nikkei and }$y_{2,t-1}${\small \ is the
S\&P 500 index return. The second and third rows are the cases where stock
return series are used. The fourth and fifth rows are the cases where
standardized residuals from the GJR-GARCH model are used. }$^{\ast }${\small %
\ indicates singificance at the 5\% level. }\pagebreak

{\small \noindent }Table 4. Cross-quantilograms at the first lag from other
markets to the US market\bigskip

\begin{center}
\begin{tabular}{llllllllll}
\hline
& $\tau _{1}$ $(=\tau _{2})$ &  & FTSE &  & DAX &  & CAC &  & Nikkei \\ 
\hline
Return & $[0,0.05]$ &  & 0.06$^{\ast }$ &  & 0.08$^{\ast }$ &  & 0.06$^{\ast
}$ &  & 0.05$^{\ast }$ \\ 
& $[0.95,1]$ &  & 0.03 &  & 0.06$^{\ast }$ &  & 0.06 &  & -0.01 \\ 
&  &  &  &  &  &  &  &  &  \\ 
Std. residual & $[0,0.05]$ &  & 0.02 &  & 0.01 &  & -0.00 &  & 0.02 \\ 
& $[0.95,1]$ &  & -0.02 &  & -0.01 &  & 0.01 &  & -0.04$^{\ast }$ \\ \hline
\end{tabular}
\end{center}

{\small \noindent Note: The table reports }${\small \hat{\rho}}_{\tau }%
{\small (1),}${\small \ a sample cross-quantilogram at the first lag, from
each stock market to the US stock market, i.e., }$y_{1,t}${\small \ is the
S\&P 500 index return and }$y_{2,t-1}${\small \ is the return series of
FTSE, DAX, CAC or Nikkei. Same as Table 3.}

\bigskip

\bigskip

{\small \noindent }Table 5. DMW test results for GJR-GARCH \bigskip

\begin{center}
\begin{tabular}{lcccccccc}
\hline
&  & FTSE &  & DAX &  & CAC &  & Nikkei \\ \hline
In-sample &  & $3.29^{\ast \ast }$ &  & $3.13^{\ast \ast }$ &  & $4.03^{\ast
\ast }$ &  & $3.59^{\ast \ast }$ \\ \hline
Out-of-sample &  & $7.67^{\ast \ast }$ &  & $10.88^{\ast \ast }$ &  & $%
12.90^{\ast \ast }$ &  & $10.01^{\ast \ast }$ \\ \hline
\end{tabular}
\end{center}

{\small \noindent Note: The table reports the DMW statistics given in (\ref%
{DMW}). The base model is the GJR-GARCH model. }$^{\ast \ast }${\small \
indicates that the null hypothesis of equal predictability between the base
model and the quantile-augmented model is rejected at the 1\% significance
level. }

\bigskip

\bigskip

{\small \noindent }Table 6. Out-of-sample forecast period and number of
forecasts \bigskip

\begin{center}
\begin{tabular}{l|l}
\hline
Index & Forecast period (number of forecasts) \\ \hline
FTSE & 2 Mar. 2006 - 31 Dec. 2007 (454 forecasts) \\ 
DAX & 1 June 2004 - 31 Dec. 2007 (891 forecasts) \\ 
CAC & 2 June 2004 - 31 Dec. 2007 (892 forecasts) \\ 
Nikkei & 29 Oct. 2004 - 31 Dec. 2007 (746 forecasts) \\ \hline
\end{tabular}
\end{center}

{\small \noindent Note: The table reports the out-of-sample forecast period
and number of forecasts for each return series. For each return series,
one-step ahead out-of-sample forecasts are produced via the rolling window
procedure with a moving window of eight years (2016 days).}

\bigskip

\bigskip

{\small \noindent }Table 7. DMW test results for GJR-GARCH with
t-distribution\bigskip

\begin{center}
\begin{tabular}{lcccccccc}
\hline
&  & FTSE &  & DAX &  & CAC &  & Nikkei \\ \hline
In-sample &  & $2.96^{\ast \ast }$ &  & $2.85^{\ast \ast }$ &  & $3.96^{\ast
\ast }$ &  & $2.91^{\ast \ast }$ \\ \hline
Out-of-sample &  & $7.28^{\ast \ast }$ &  & $8.71^{\ast \ast }$ &  & $%
12.56^{\ast \ast }$ &  & $8.98^{\ast \ast }$ \\ \hline
\end{tabular}
\end{center}

{\small \noindent Note: The base model is the GJR-GARCH model with }$t$%
{\small -distribution. Same as Table 5. }

\pagebreak

{\small \noindent }Table 8. DMW test results for HEAVY

\begin{center}
\begin{tabular}{lcccccccc}
\hline
&  & FTSE &  & DAX &  & CAC &  & Nikkei \\ \hline
In-sample &  & $3.78^{\ast \ast }$ &  & $3.20^{\ast \ast }$ &  & $4.86^{\ast
\ast }$ &  & $2.89^{\ast \ast }$ \\ \hline
Out-of-sample &  & $7.22^{\ast \ast }$ &  & $8.19^{\ast \ast }$ &  & $%
10.84^{\ast \ast }$ &  & $8.43^{\ast \ast }$ \\ \hline
\end{tabular}
\end{center}

{\small \noindent Note: The base model is the HEAVY-r model by Shephard and
Sheppard (2010). Same as Table 5. }

\bigskip

\bigskip

\begin{center}
\FRAME{itbpF}{2.6024in}{1.2631in}{0in}{}{}{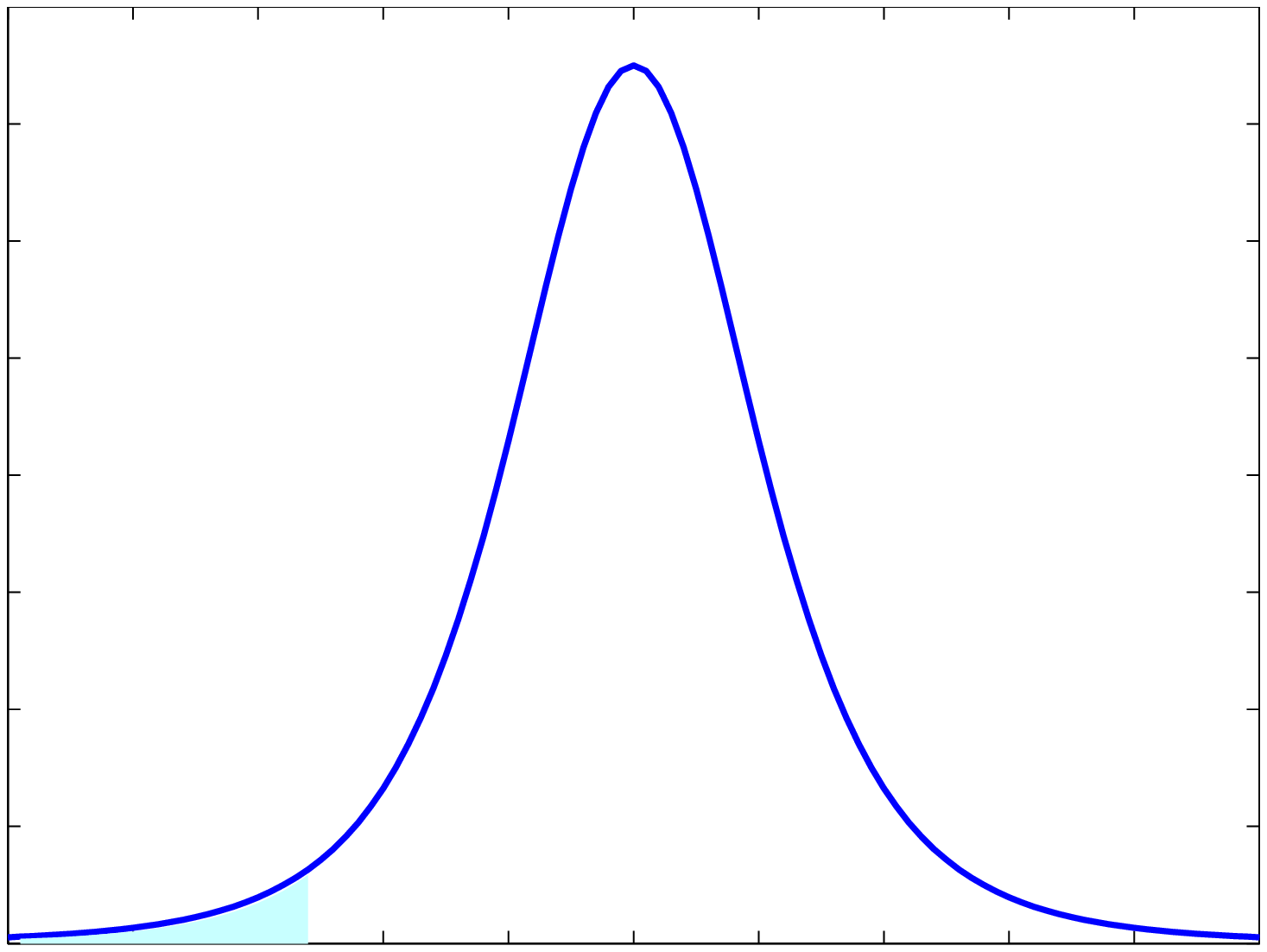}{\special{language
"Scientific Word";type "GRAPHIC";display "PICT";valid_file "F";width
2.6024in;height 1.2631in;depth 0in;original-width 5.9836in;original-height
4.4988in;cropleft "0";croptop "1";cropright "1";cropbottom "0";filename
'../Codes/f1.eps';file-properties "XNPEU";}} \ \FRAME{itbpF}{2.6024in}{%
1.2631in}{0in}{}{}{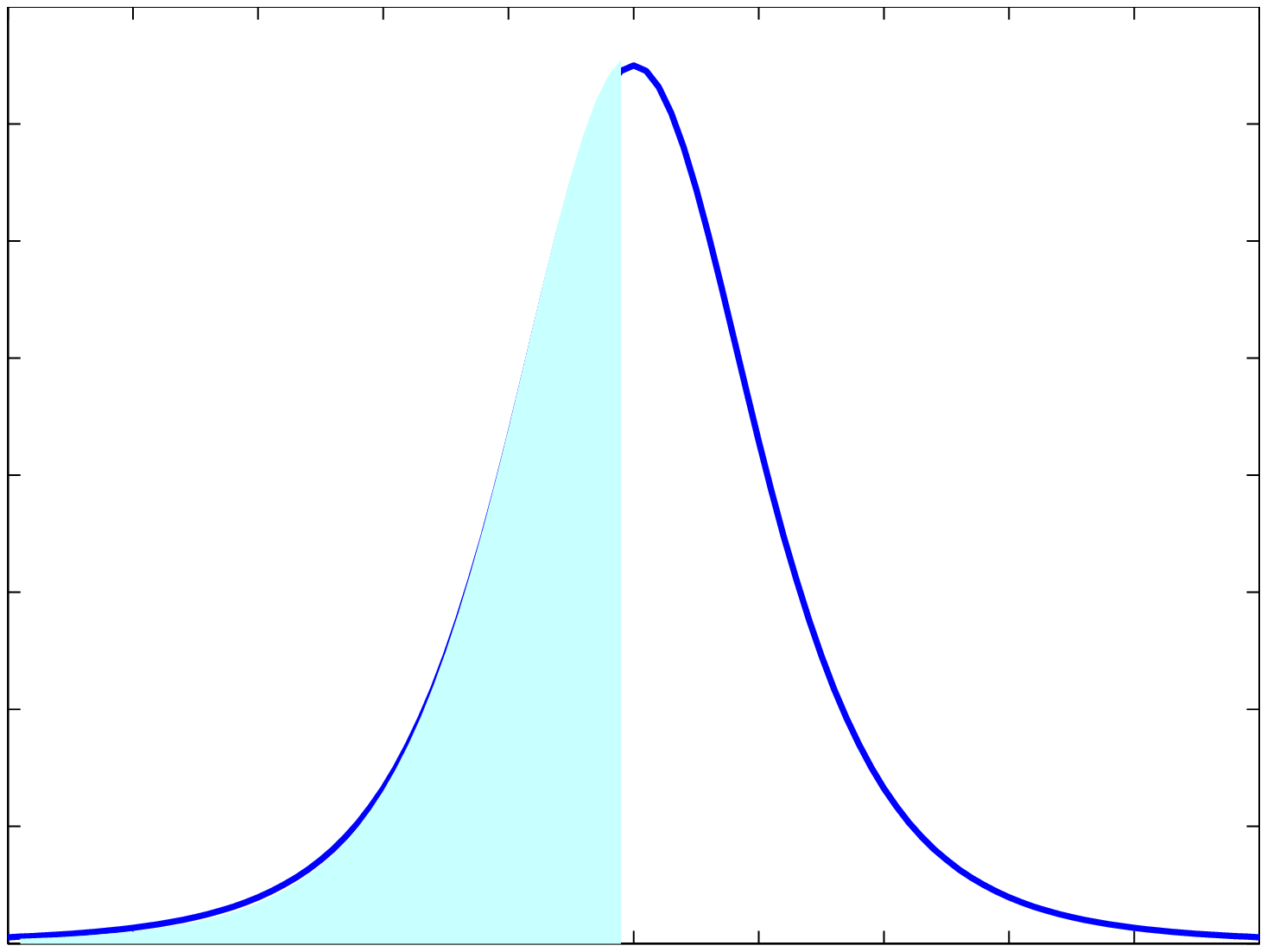}{\special{language "Scientific Word";type
"GRAPHIC";display "PICT";valid_file "F";width 2.6024in;height 1.2631in;depth
0in;original-width 5.9836in;original-height 4.4988in;cropleft "0";croptop
"1";cropright "1";cropbottom "0";filename '../Codes/f2.eps';file-properties
"XNPEU";}}
\end{center}

{\small \noindent Figure 1. Event }${\small \{y}_{i,t}{\small <q}_{i,t}%
{\small (\tau }_{i}{\small )\}.}${\small \ The left figure describes an
event }${\small \{y}_{1,t}{\small <q}_{1,t}{\small (\tau }_{1}{\small )\}}$%
{\small \ for }${\small \tau }_{1}{\small =0.05}${\small \ and the right
figure provides an event }${\small \{y}_{2,t-k}{\small <q}_{2,t-k}{\small %
(\tau }_{2}{\small )\}}${\small \ for }${\small \tau }_{2}{\small =0.5.}$

\bigskip

\begin{center}
\FRAME{itbpF}{2.6024in}{1.2639in}{0in}{}{}{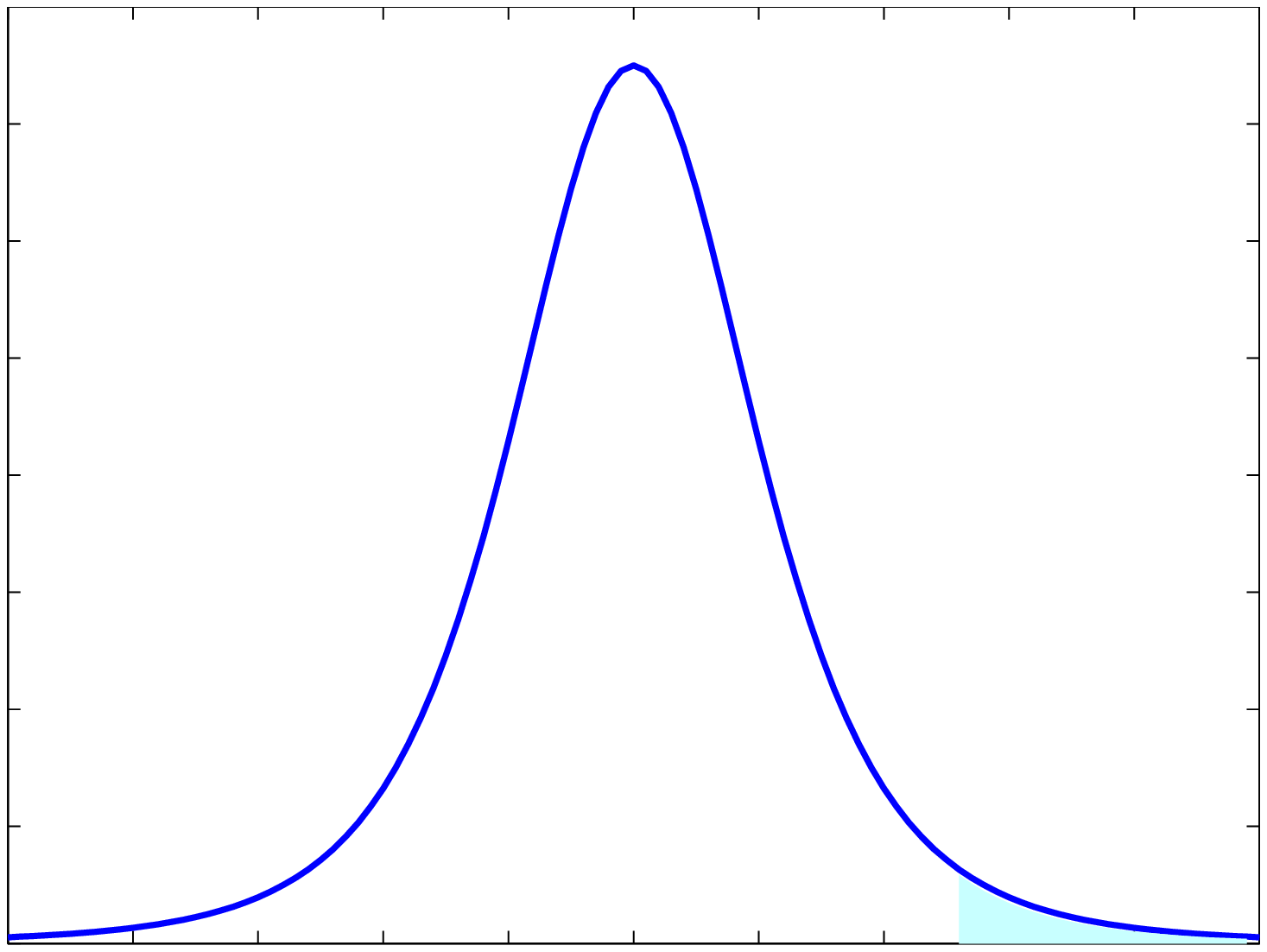}{\special{language
"Scientific Word";type "GRAPHIC";display "PICT";valid_file "F";width
2.6024in;height 1.2639in;depth 0in;original-width 5.9836in;original-height
4.4988in;cropleft "0";croptop "1";cropright "1";cropbottom "0";filename
'../Codes/f3.eps';file-properties "XNPEU";}} \ \FRAME{itbpF}{2.6024in}{%
1.2639in}{0in}{}{}{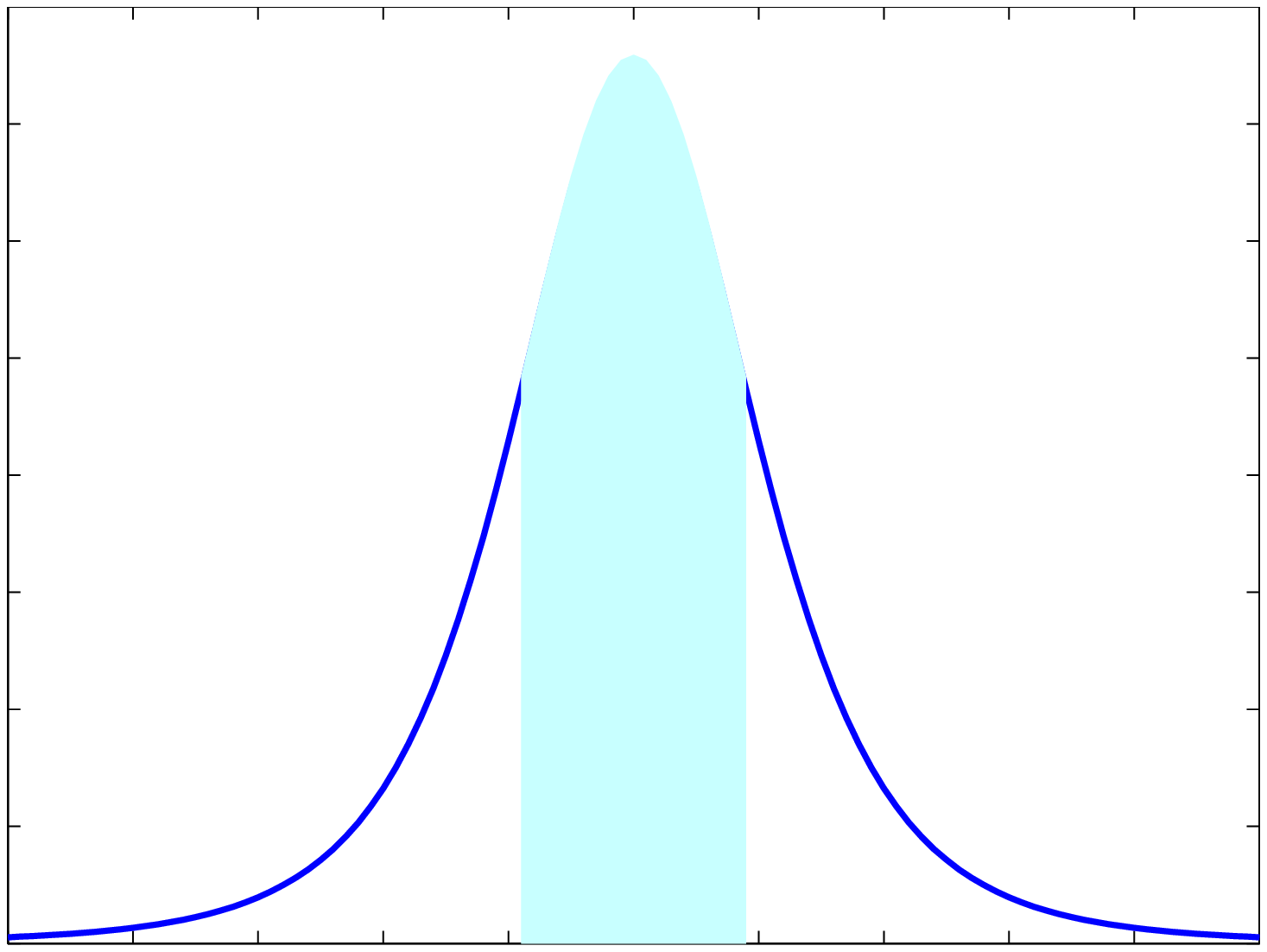}{\special{language "Scientific Word";type
"GRAPHIC";display "PICT";valid_file "F";width 2.6024in;height 1.2639in;depth
0in;original-width 5.9836in;original-height 4.4988in;cropleft "0";croptop
"1";cropright "1";cropbottom "0";filename '../Codes/f4.eps';file-properties
"XNPEU";}}

\FRAME{itbpF}{2.6024in}{1.2639in}{0in}{}{}{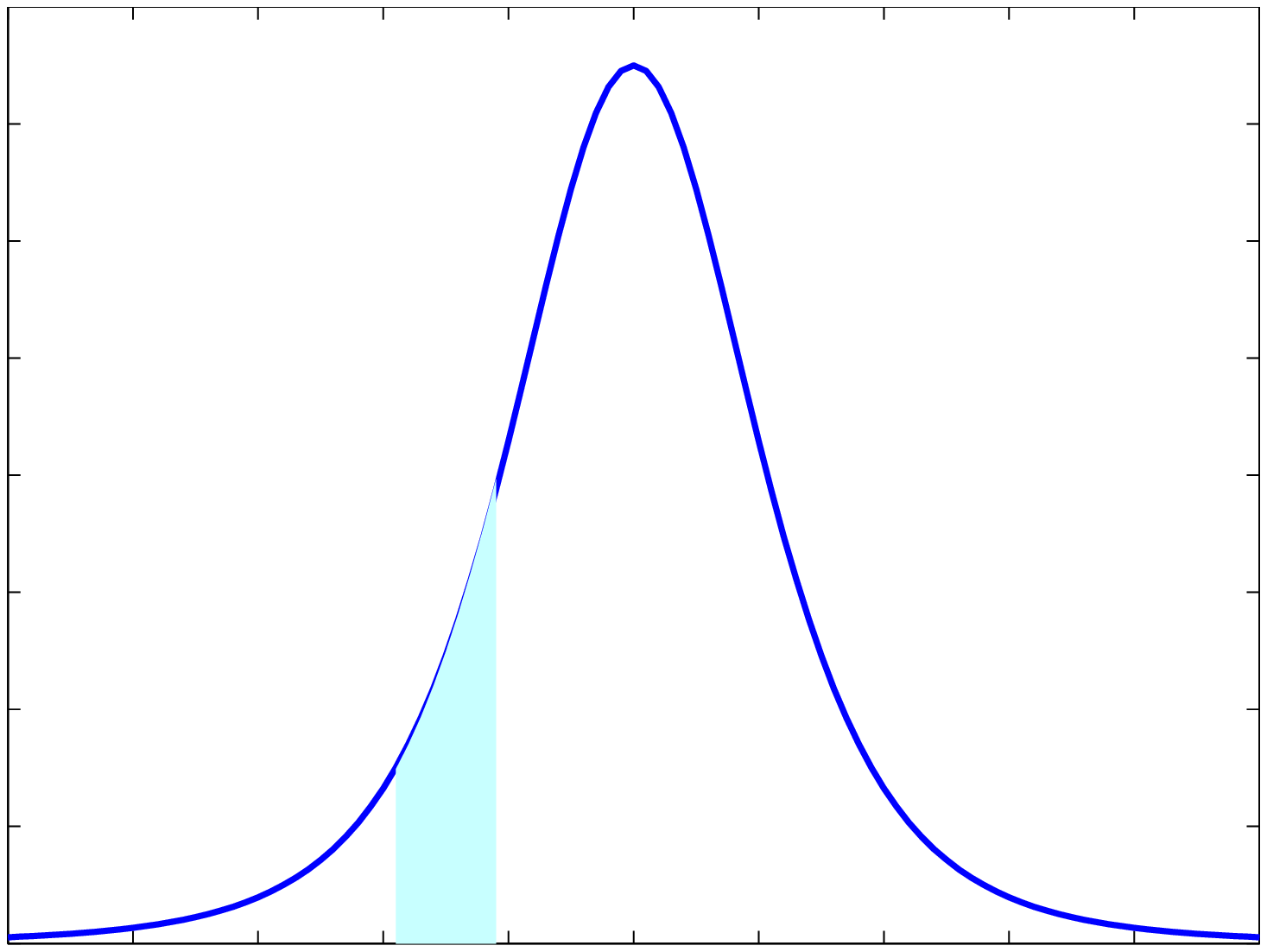}{\special{language
"Scientific Word";type "GRAPHIC";display "PICT";valid_file "F";width
2.6024in;height 1.2639in;depth 0in;original-width 5.9836in;original-height
4.4988in;cropleft "0";croptop "1";cropright "1";cropbottom "0";filename
'../Codes/f5.eps';file-properties "XNPEU";}} \ \FRAME{itbpF}{2.6024in}{%
1.2639in}{0in}{}{}{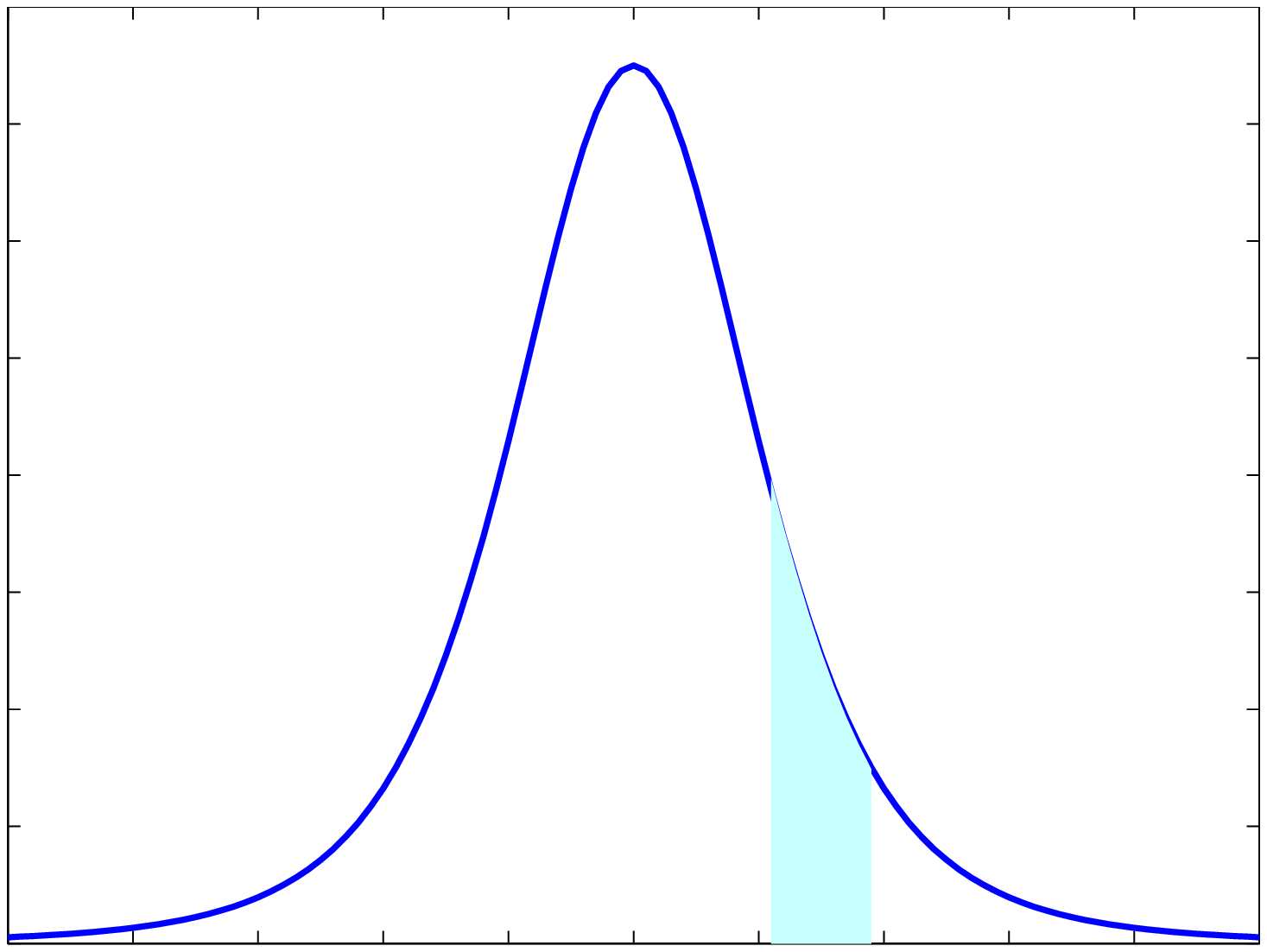}{\special{language "Scientific Word";type
"GRAPHIC";display "PICT";valid_file "F";width 2.6024in;height 1.2639in;depth
0in;original-width 5.9836in;original-height 4.4988in;cropleft "0";croptop
"1";cropright "1";cropbottom "0";filename '../Codes/f6.eps';file-properties
"XNPEU";}}
\end{center}

{\small \noindent Figure 2. Event }${\small \{q}_{i,t}{\small (\tau }_{i}^{l}%
{\small )<y}_{i,t}{\small <q}_{i,t}{\small (\tau }_{i}^{h}{\small )\}.}$%
{\small \ The figures describe various events }${\small \{q}_{i,t}{\small %
(\tau }_{i}^{l}{\small )<y}_{i,t}{\small <q}_{i,t}{\small (\tau }_{i}^{h}%
{\small )\}}${\small \ for different quantiles for }$\tau _{i}^{l}${\small \
and }$\tau _{i}^{h}.${\small \ The top left figure provides a right-tail
event and the top right figure gives a mid-range event. The bottom figures
present events for the left and right shoulders of the distribution.}

\pagebreak

\begin{center}
\FRAME{itbpF}{6.263in}{3.2499in}{0in}{}{}{autous.eps}{\special{language
"Scientific Word";type "GRAPHIC";maintain-aspect-ratio TRUE;display
"PICT";valid_file "F";width 6.263in;height 3.2499in;depth 0in;original-width
6.3543in;original-height 3.2827in;cropleft "0";croptop "1";cropright
"1";cropbottom "0";filename '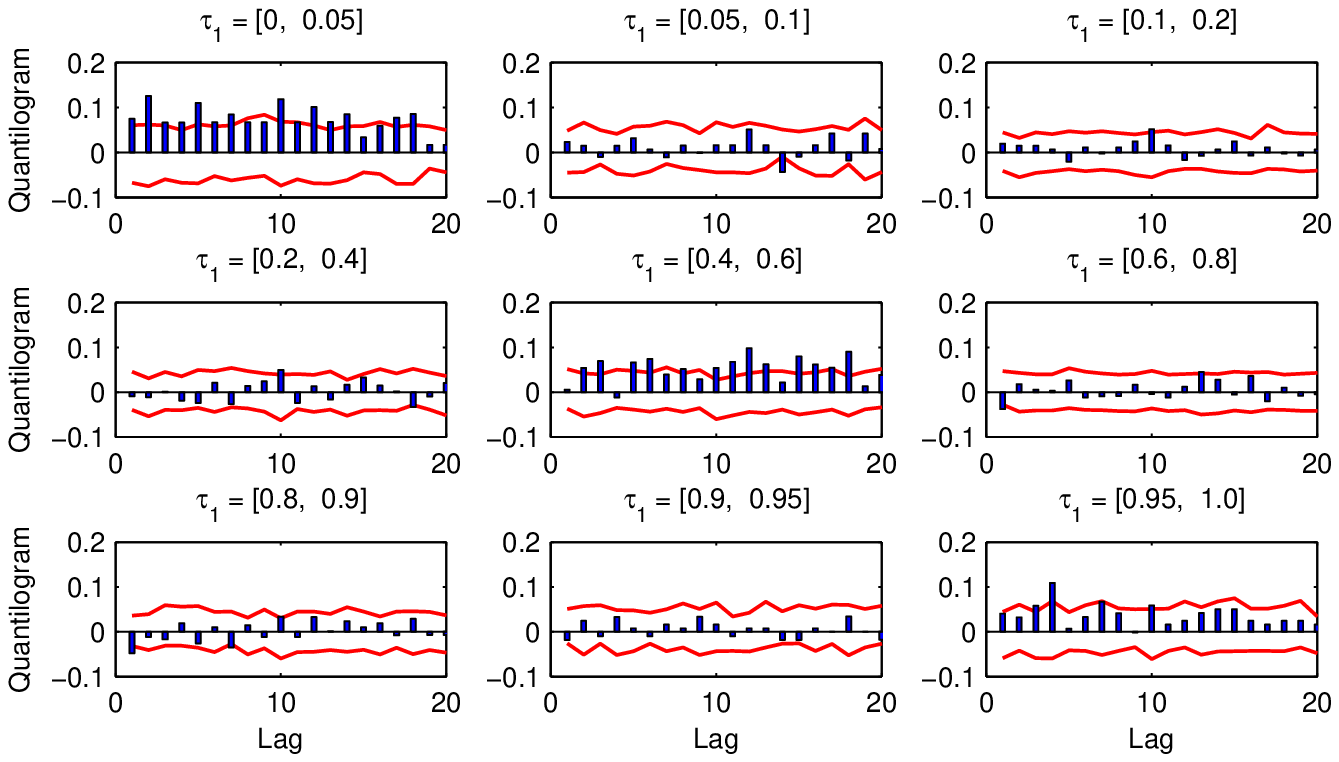';file-properties "XNPEU";}}
\end{center}

{\small \noindent Figure 3(a). [US] Auto-quantilogram }${\small \hat{\rho}}%
_{\tau }{\small (k)}${\small \ of the S\&P 500 index return series. }$\tau
_{1}${\small \ is the quantile range. Bar graphs describe sample
cross-quantilograms and lines are the 95\% bootstrap confidence intervals
centered at zero.}

\begin{center}
\FRAME{itbpF}{6.263in}{3.2499in}{0in}{}{}{autous_q.eps}{\special{language
"Scientific Word";type "GRAPHIC";maintain-aspect-ratio TRUE;display
"PICT";valid_file "F";width 6.263in;height 3.2499in;depth 0in;original-width
6.3543in;original-height 3.2827in;cropleft "0";croptop "1";cropright
"1";cropbottom "0";filename '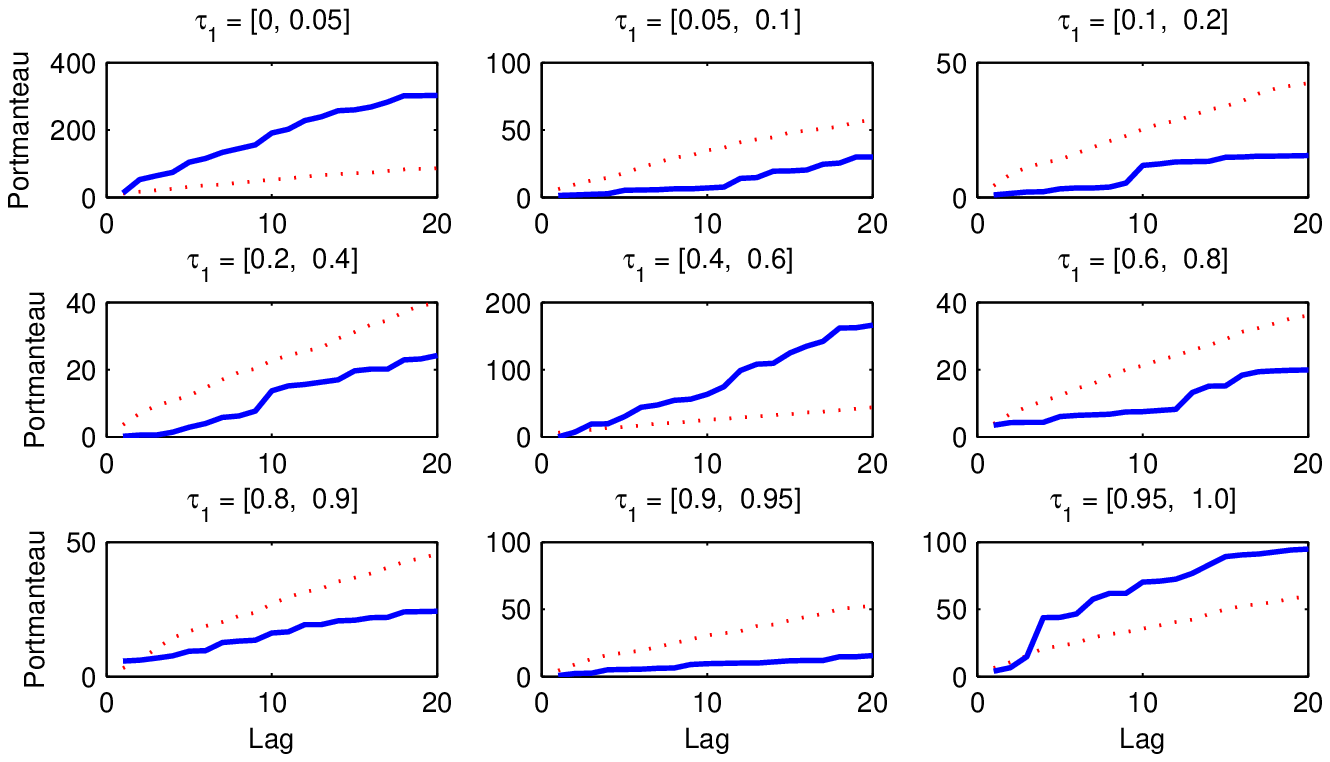';file-properties "XNPEU";}}
\end{center}

{\small \noindent Figure 3(b). [US] Box-Ljung test statistic }${\small \hat{Q%
}}_{\tau }^{(p)}${\small \ for each lag }${\small p}${\small \ using }$%
{\small \hat{\rho}}_{\tau }{\small (k)}${\small . Same as Figure 1(a). The
dashed lines are the 95\% bootstrap confidence intervals centered at
zero.\pagebreak }

\begin{center}
\FRAME{itbpF}{6.263in}{3.2499in}{0in}{}{}{autouk.eps}{\special{language
"Scientific Word";type "GRAPHIC";maintain-aspect-ratio TRUE;display
"PICT";valid_file "F";width 6.263in;height 3.2499in;depth 0in;original-width
6.3543in;original-height 3.2827in;cropleft "0";croptop "1";cropright
"1";cropbottom "0";filename '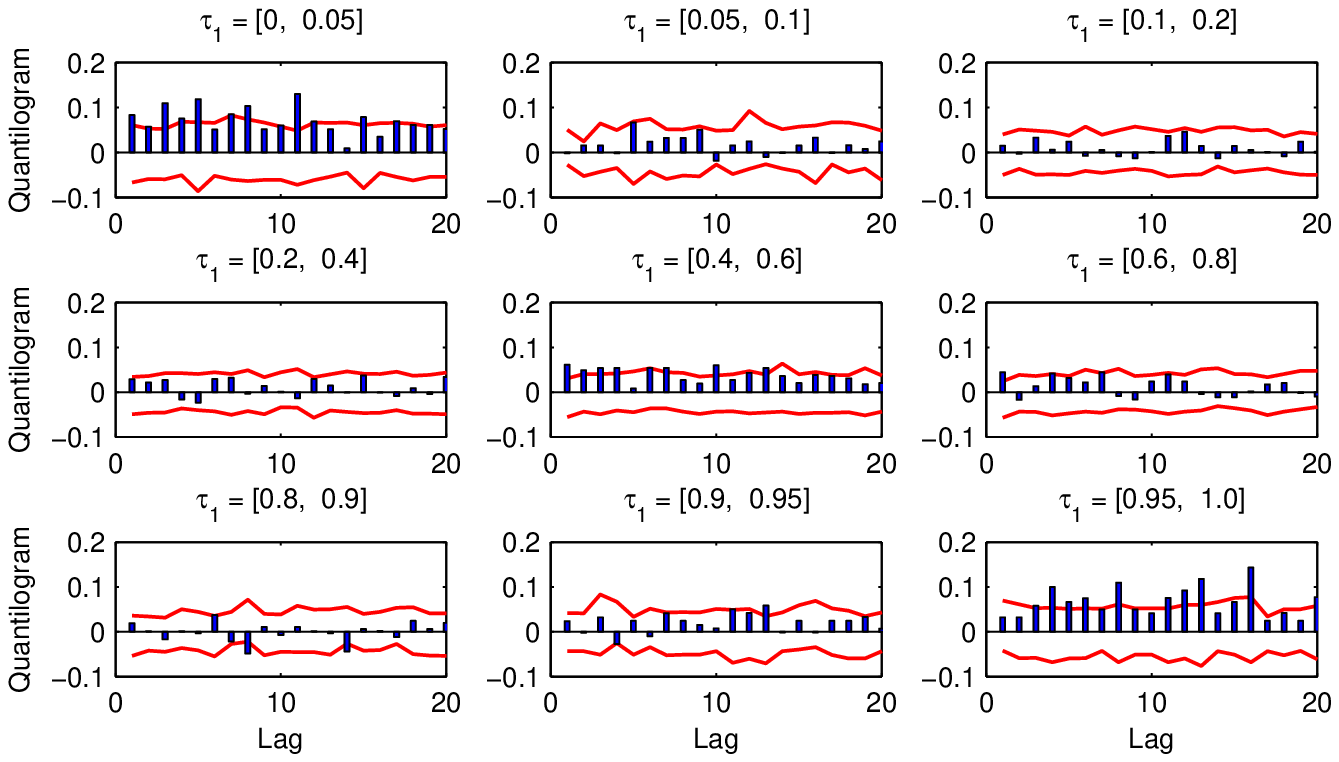';file-properties "XNPEU";}}
\end{center}

{\small \noindent Figure 4(a). [UK] Auto-quantilogram }${\small \hat{\rho}}%
_{\tau }{\small (k)}${\small \ of the FTSE index return series.\ Same as
Figure 1(a).}

\begin{center}
\FRAME{itbpF}{6.263in}{3.2499in}{0in}{}{}{autouk_q.eps}{\special{language
"Scientific Word";type "GRAPHIC";maintain-aspect-ratio TRUE;display
"PICT";valid_file "F";width 6.263in;height 3.2499in;depth 0in;original-width
6.3543in;original-height 3.2827in;cropleft "0";croptop "1";cropright
"1";cropbottom "0";filename '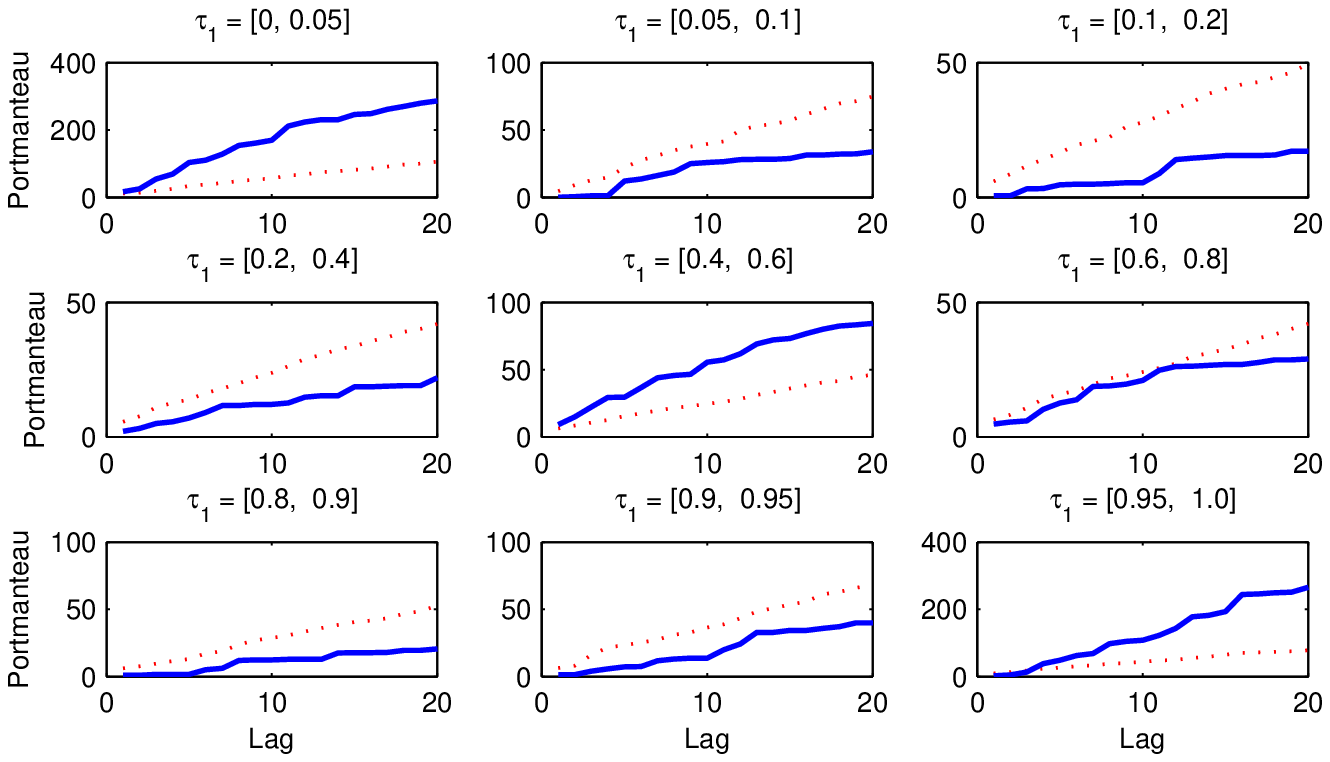';file-properties "XNPEU";}}
\end{center}

{\small \noindent Figure 4(b). [UK] Box-Ljung test statistic }${\small \hat{Q%
}}_{\tau }^{(p)}${\small \ for each lag }${\small p}${\small \ using }$%
{\small \hat{\rho}}_{\tau }{\small (k)}${\small . Same as Figure
1(b).\pagebreak\ }

\begin{center}
\FRAME{itbpF}{6.263in}{3.2499in}{0in}{}{}{ustouk.eps}{\special{language
"Scientific Word";type "GRAPHIC";maintain-aspect-ratio TRUE;display
"PICT";valid_file "F";width 6.263in;height 3.2499in;depth 0in;original-width
6.3543in;original-height 3.2827in;cropleft "0";croptop "1";cropright
"1";cropbottom "0";filename '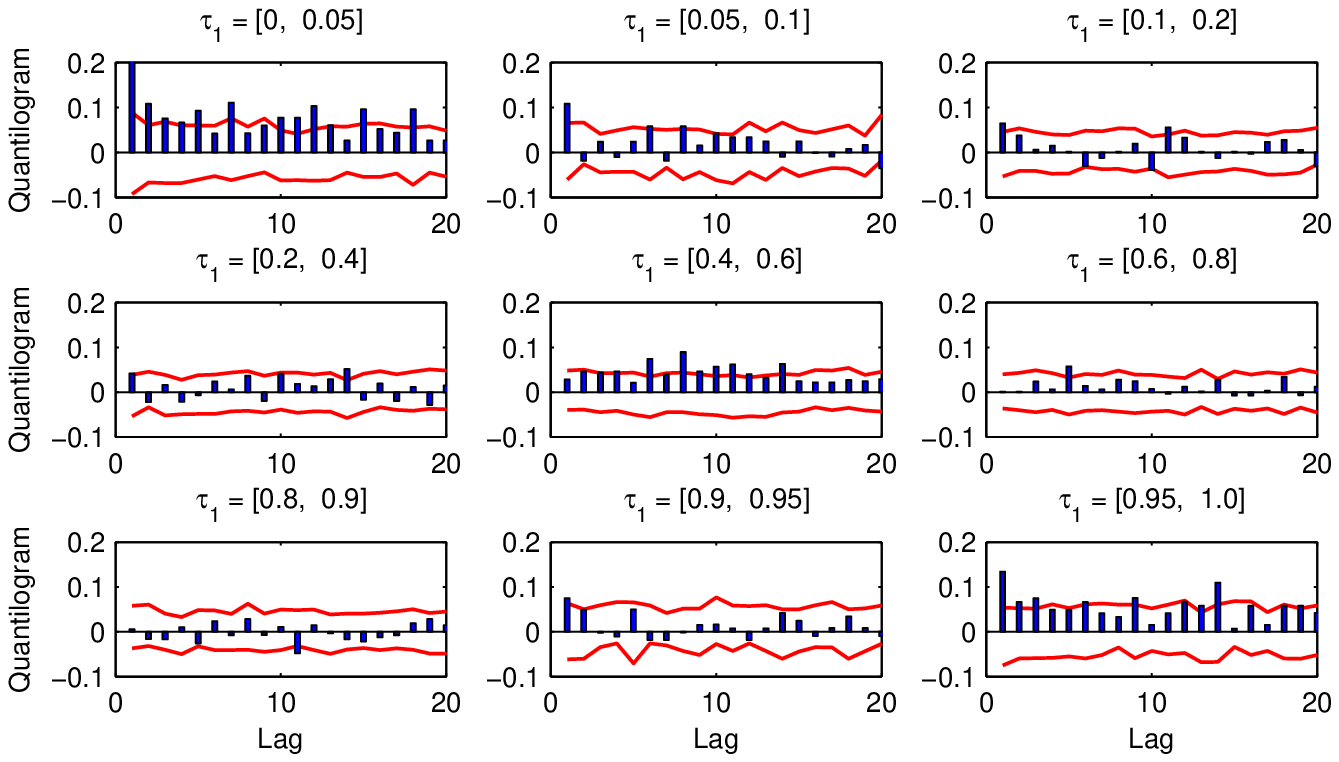';file-properties "XNPEU";}}
\end{center}

{\small \noindent Figure 5(a). [US to UK] Cross-quantilogram }${\small \hat{%
\rho}}_{\tau }{\small (k)}${\small \ to detect directional predictability
from US to UK. }$\tau _{1}${\small =}$\tau _{2}${\small . Same as Figure
1(a).}

\begin{center}
\FRAME{itbpF}{6.263in}{3.2499in}{0in}{}{}{ustouk_q.eps}{\special{language
"Scientific Word";type "GRAPHIC";maintain-aspect-ratio TRUE;display
"PICT";valid_file "F";width 6.263in;height 3.2499in;depth 0in;original-width
6.3543in;original-height 3.2827in;cropleft "0";croptop "1";cropright
"1";cropbottom "0";filename '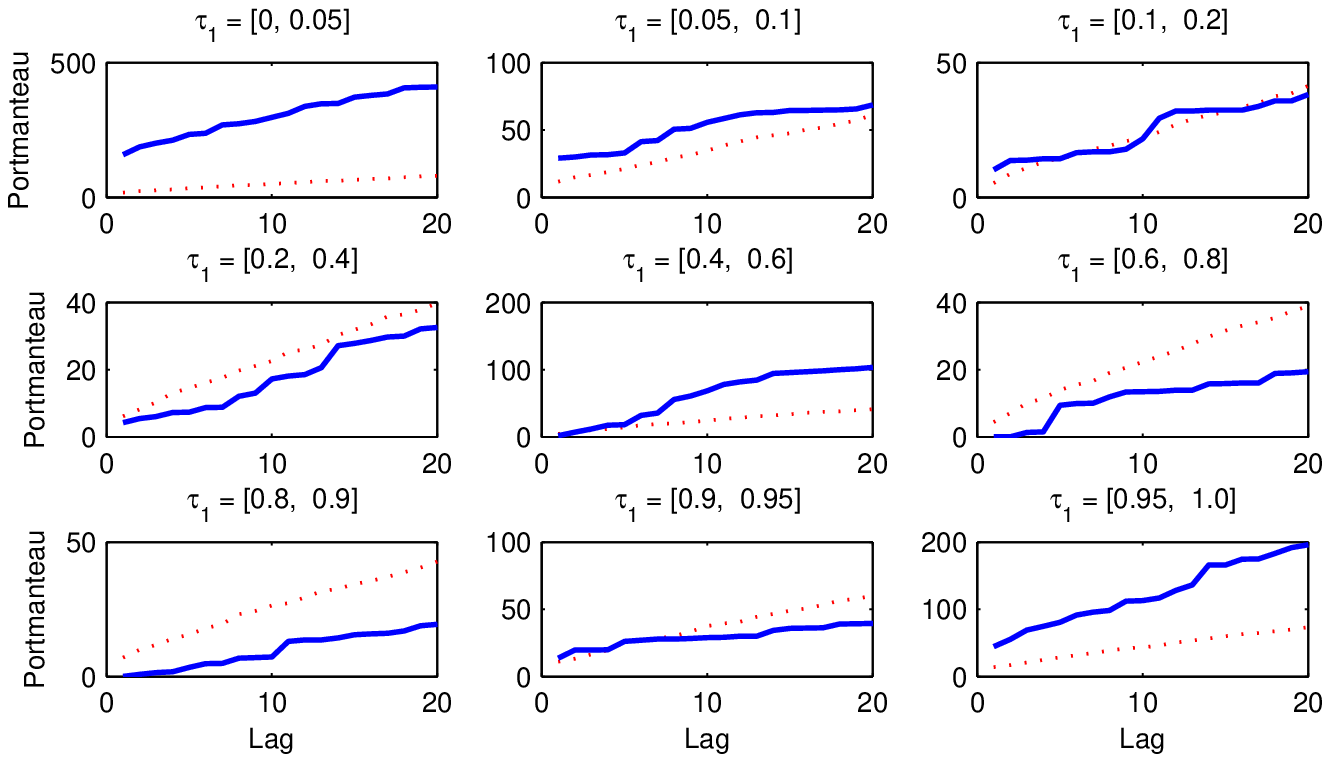';file-properties "XNPEU";}}
\end{center}

{\small \noindent Figure 5(b). [US to UK] Box-Ljung test statistic }${\small 
\hat{Q}}_{\tau }^{(p)}${\small \ for each lag }${\small p}${\small \ using }$%
{\small \hat{\rho}}_{\tau }{\small (k)}${\small . Same as Figure
1(b).\pagebreak\ }

\begin{center}
\FRAME{itbpF}{6.263in}{3.2499in}{0in}{}{}{tousfromuk.eps}{\special{language
"Scientific Word";type "GRAPHIC";maintain-aspect-ratio TRUE;display
"PICT";valid_file "F";width 6.263in;height 3.2499in;depth 0in;original-width
6.3543in;original-height 3.2827in;cropleft "0";croptop "1";cropright
"1";cropbottom "0";filename '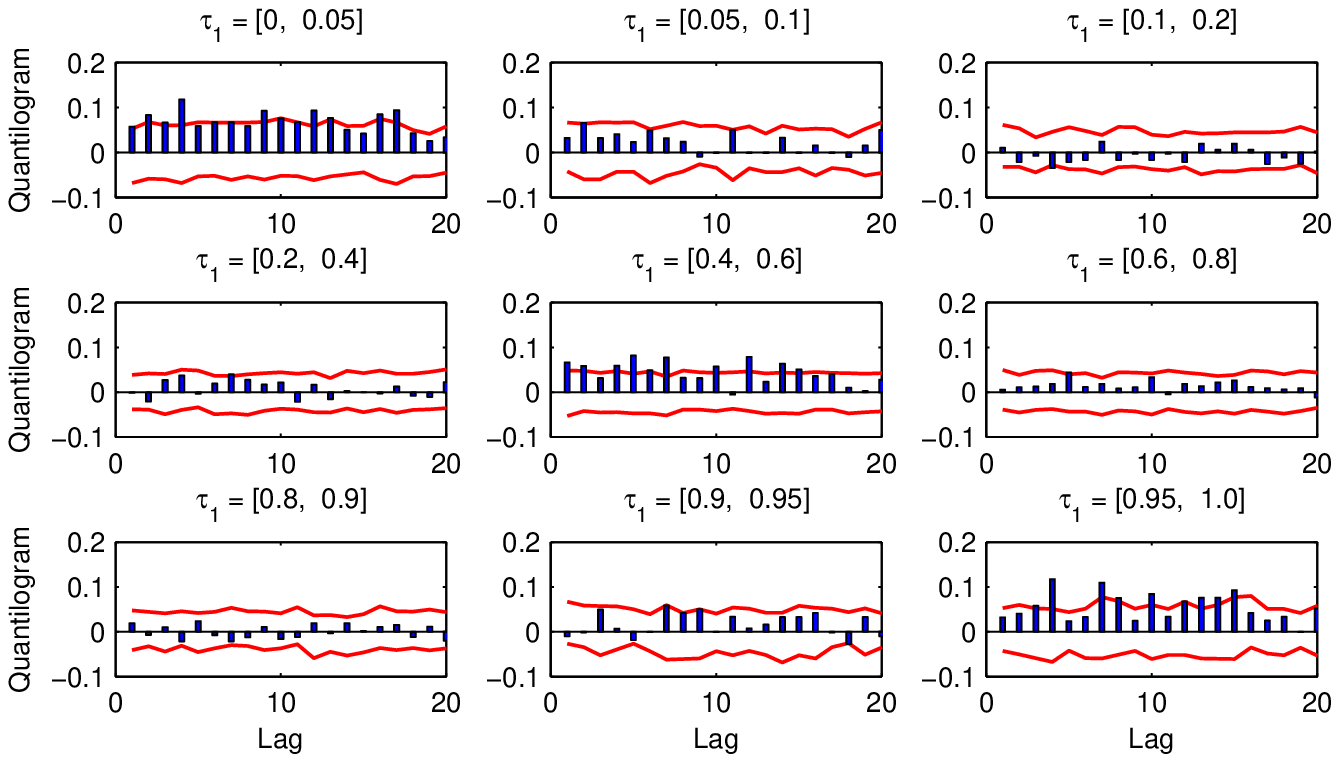';file-properties "XNPEU";}}
\end{center}

{\small \noindent Figure 6(a). [UK to US] Cross-quantilogram }${\small \hat{%
\rho}}_{\tau }{\small (k)}$\ {\small to detect directional predictability
from UK to US }$\tau _{1}${\small =}$\tau _{2}${\small . Same as Figure 1(a).%
}

\begin{center}
\FRAME{itbpF}{6.263in}{3.2499in}{0in}{}{}{tousfromuk_q.eps}{\special%
{language "Scientific Word";type "GRAPHIC";maintain-aspect-ratio
TRUE;display "PICT";valid_file "F";width 6.263in;height 3.2499in;depth
0in;original-width 6.3543in;original-height 3.2827in;cropleft "0";croptop
"1";cropright "1";cropbottom "0";filename '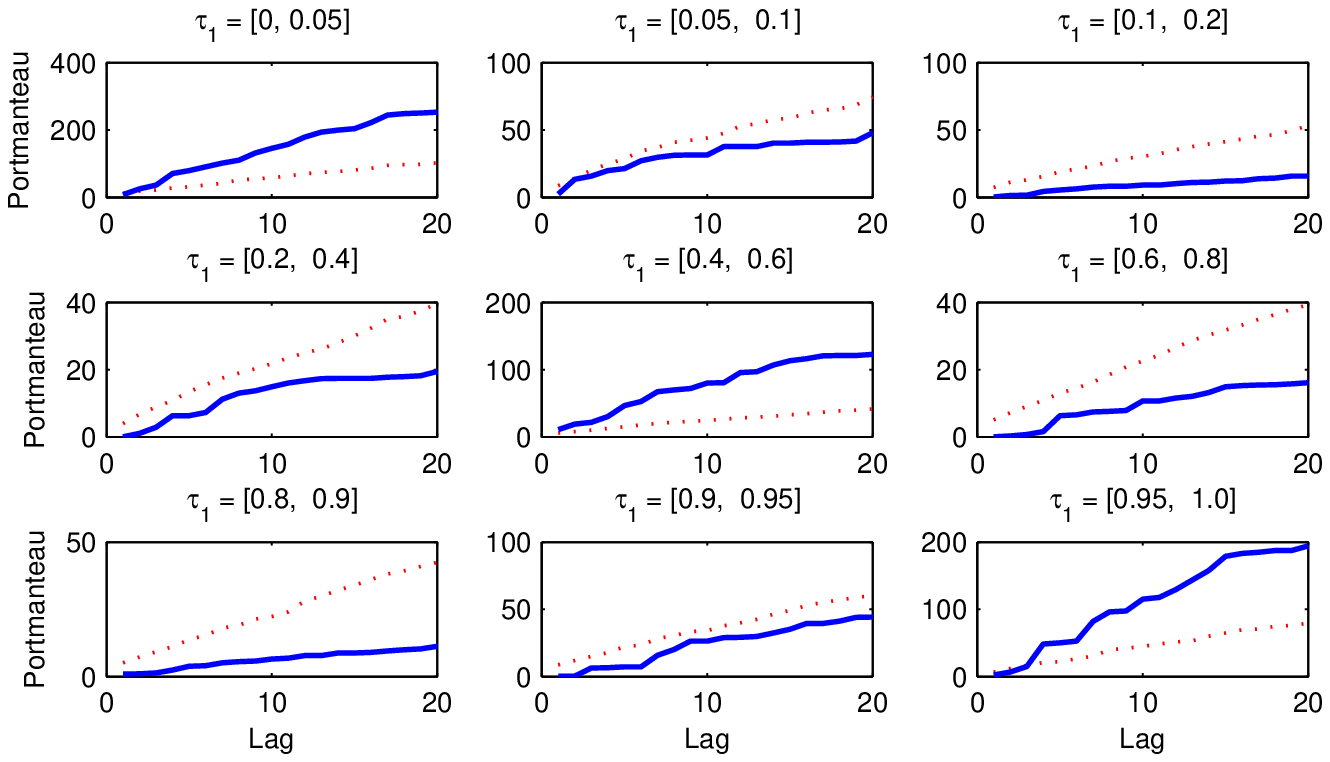';file-properties
"XNPEU";}}
\end{center}

{\small \noindent Figure 6(b). [UK to US] Box-Ljung test statistic }${\small 
\hat{Q}}_{\tau }^{(p)}${\small \ for each lag }${\small p}${\small \ using }$%
{\small \hat{\rho}}_{\tau }{\small (k)}${\small . Same as Figure
4(b).\pagebreak\ }

\begin{center}
\FRAME{itbpF}{6.263in}{3.2499in}{0in}{}{}{sdgjrautous.eps}{\special{language
"Scientific Word";type "GRAPHIC";maintain-aspect-ratio TRUE;display
"PICT";valid_file "F";width 6.263in;height 3.2499in;depth 0in;original-width
6.3543in;original-height 3.2827in;cropleft "0";croptop "1";cropright
"1";cropbottom "0";filename '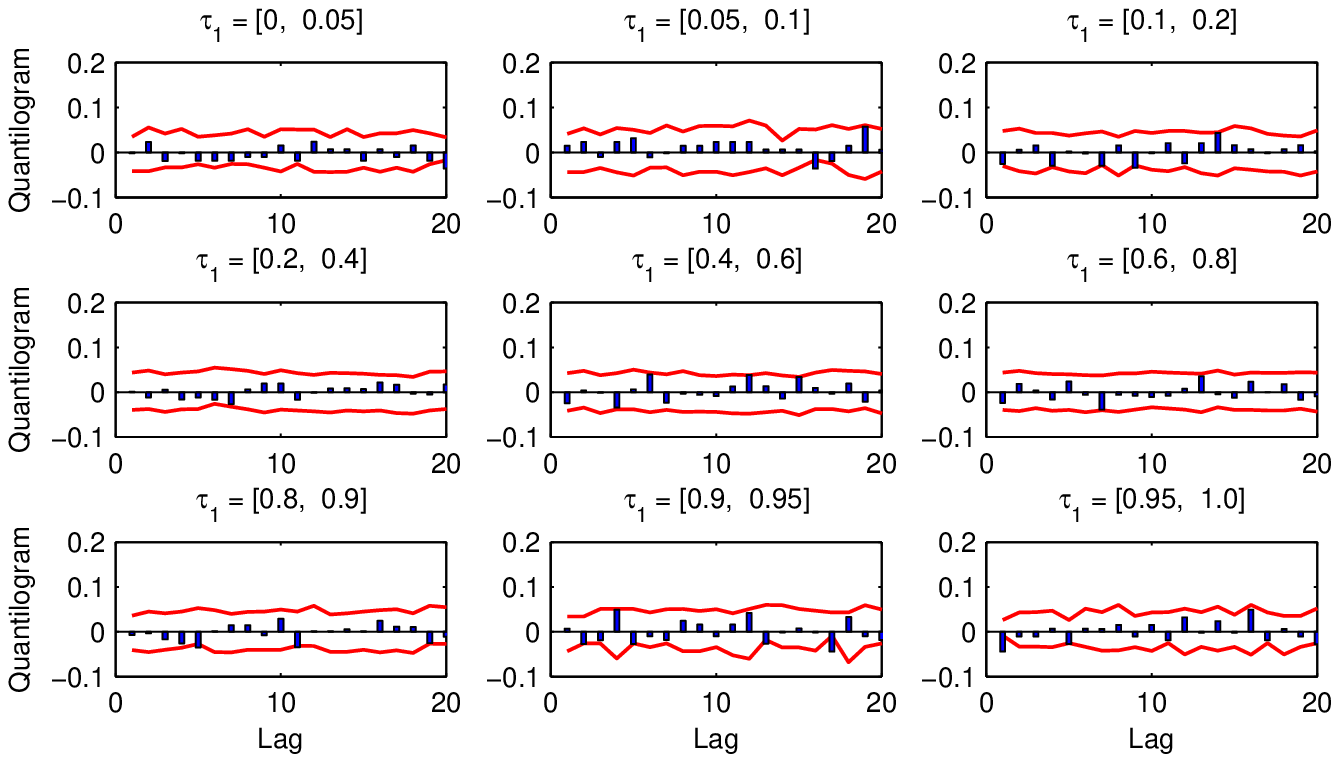';file-properties "XNPEU";}}
\end{center}

{\small \noindent Figure 7(a). [US, std. residual] Auto-quantilogram }$%
{\small \hat{\rho}}_{\tau }{\small (k)}${\small \ of the S\&P 500 index
return series using the standardized residual from the GJR-GARCH model. }$%
\tau _{1}${\small =}$\tau _{2}${\small . Same as Figure 1(a). }

\begin{center}
\FRAME{itbpF}{6.263in}{3.2499in}{0in}{}{}{sdgjrautous_q.eps}{\special%
{language "Scientific Word";type "GRAPHIC";maintain-aspect-ratio
TRUE;display "PICT";valid_file "F";width 6.263in;height 3.2499in;depth
0in;original-width 6.3543in;original-height 3.2827in;cropleft "0";croptop
"1";cropright "1";cropbottom "0";filename
'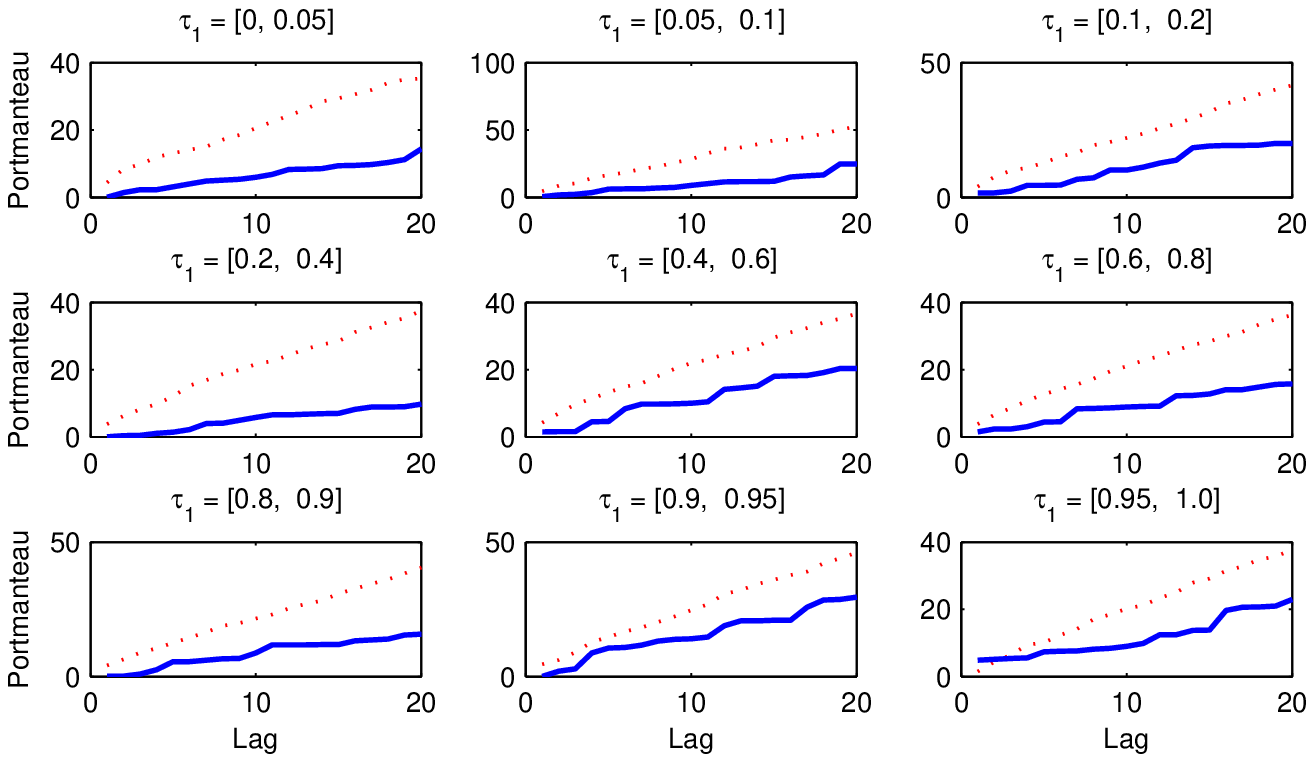';file-properties "XNPEU";}}
\end{center}

{\small \noindent Figure 7(b). [US, std. residual] Box-Ljung test statistic }%
${\small \hat{Q}}_{\tau }^{(p)}${\small \ for each lag }${\small p}${\small %
\ using }${\small \hat{\rho}}_{\tau }{\small (k)}${\small . Same as Figure
1(b).\pagebreak\ }

\begin{center}
\FRAME{itbpF}{6.263in}{3.2499in}{0in}{}{}{sdgjrautouk.eps}{\special{language
"Scientific Word";type "GRAPHIC";maintain-aspect-ratio TRUE;display
"PICT";valid_file "F";width 6.263in;height 3.2499in;depth 0in;original-width
6.3543in;original-height 3.2827in;cropleft "0";croptop "1";cropright
"1";cropbottom "0";filename '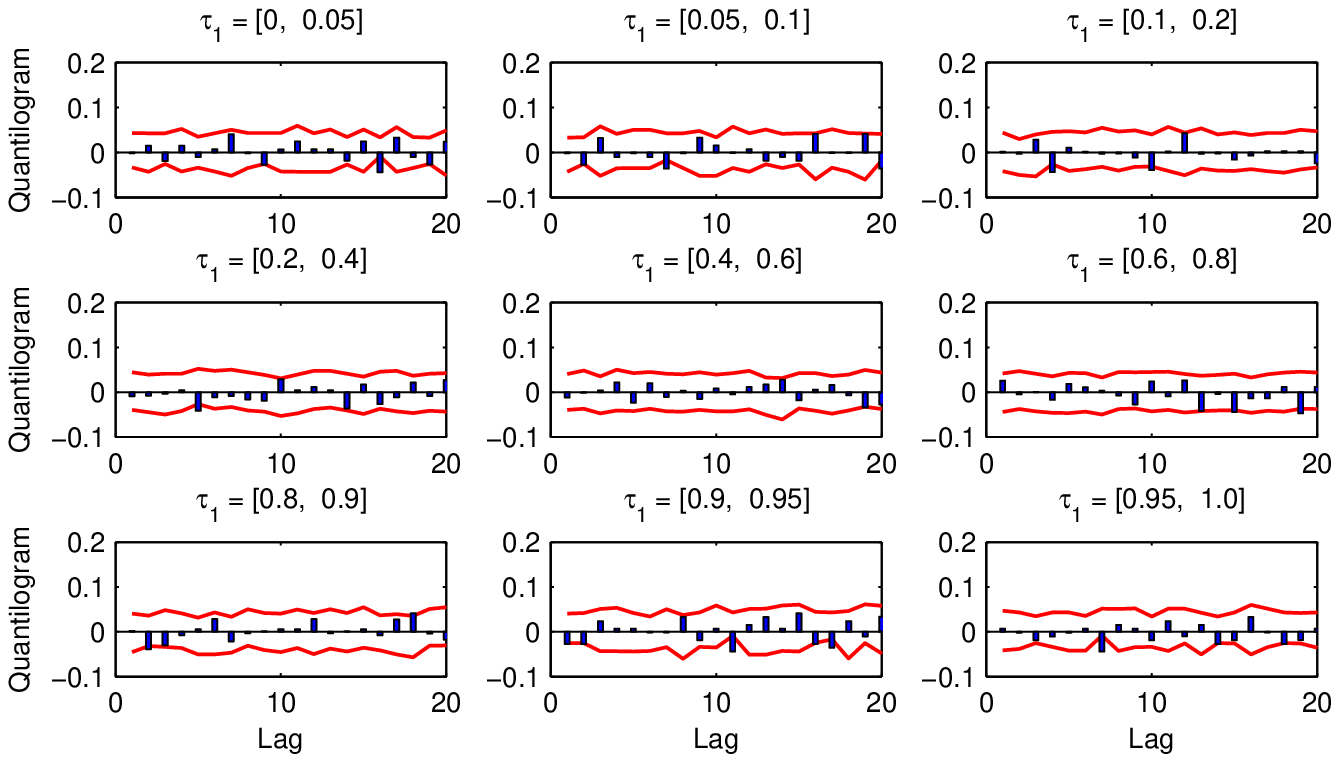';file-properties "XNPEU";}}
\end{center}

{\small \noindent Figure 8(a). [UK, std. residual] Auto-quantilogram }$%
{\small \hat{\rho}}_{\tau }{\small (k)}${\small \ of the FTSE index return
series using the standardized residual from the GJR-GARCH model. }$\tau _{1}$%
{\small =}$\tau _{2}${\small . Same as Figure 1(a).}

\begin{center}
\FRAME{itbpF}{6.263in}{3.2499in}{0in}{}{}{sdgjrautouk_q.eps}{\special%
{language "Scientific Word";type "GRAPHIC";maintain-aspect-ratio
TRUE;display "PICT";valid_file "F";width 6.263in;height 3.2499in;depth
0in;original-width 6.3543in;original-height 3.2827in;cropleft "0";croptop
"1";cropright "1";cropbottom "0";filename
'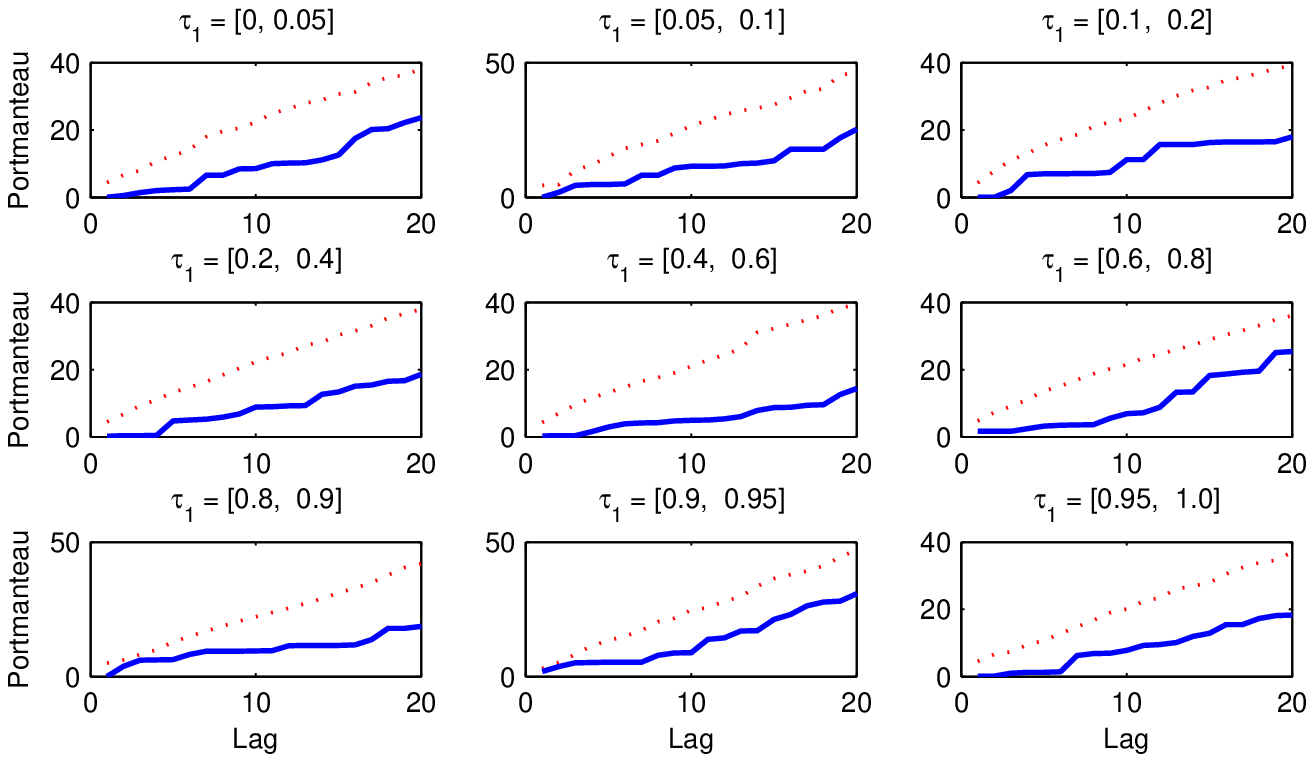';file-properties "XNPEU";}}
\end{center}

{\small \noindent Figure 8(b). [UK, std. residual] Box-Ljung test statistic }%
${\small \hat{Q}}_{\tau }^{(p)}${\small \ for each lag }${\small p}${\small %
\ using }${\small \hat{\rho}}_{\tau }{\small (k)}${\small . Same as Figure
1(b).\pagebreak\ }

\begin{center}
\FRAME{itbpF}{6.263in}{3.2499in}{0in}{}{}{sdgjr_ustouk.eps}{\special%
{language "Scientific Word";type "GRAPHIC";maintain-aspect-ratio
TRUE;display "PICT";valid_file "F";width 6.263in;height 3.2499in;depth
0in;original-width 6.3543in;original-height 3.2827in;cropleft "0";croptop
"1";cropright "1";cropbottom "0";filename '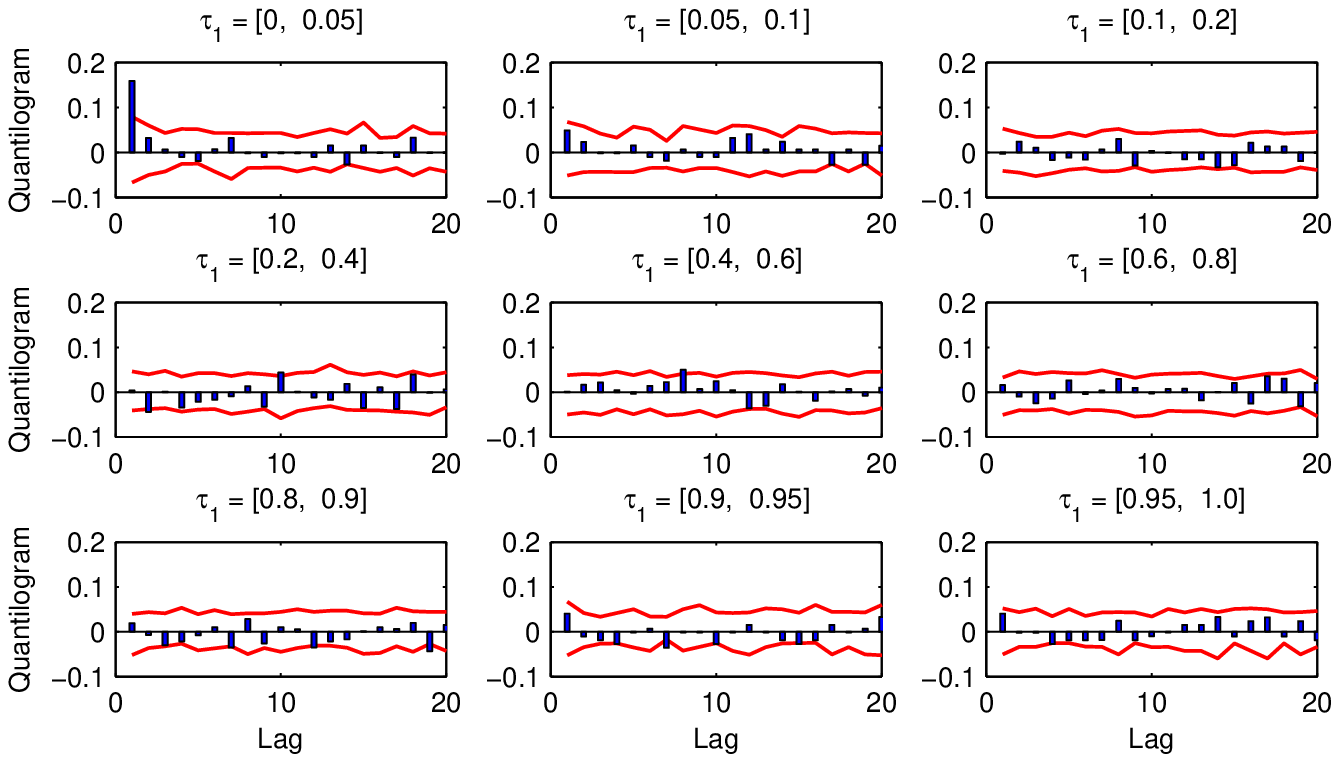';file-properties
"XNPEU";}}
\end{center}

{\small \noindent Figure 9(a). [US to UK, std. residual] Cross-quantilogram }%
${\small \hat{\rho}}_{\tau }{\small (k)}${\small \ to detect directional
predictability from US to UK using the standardized residual from the
GJR-GARCH model. }$\tau _{1}${\small =}$\tau _{2}${\small . Same as Figure
1(a).}

\begin{center}
\FRAME{itbpF}{6.263in}{3.2499in}{0in}{}{}{sdgjr_ustouk_q.eps}{\special%
{language "Scientific Word";type "GRAPHIC";maintain-aspect-ratio
TRUE;display "PICT";valid_file "F";width 6.263in;height 3.2499in;depth
0in;original-width 6.3543in;original-height 3.2827in;cropleft "0";croptop
"1";cropright "1";cropbottom "0";filename
'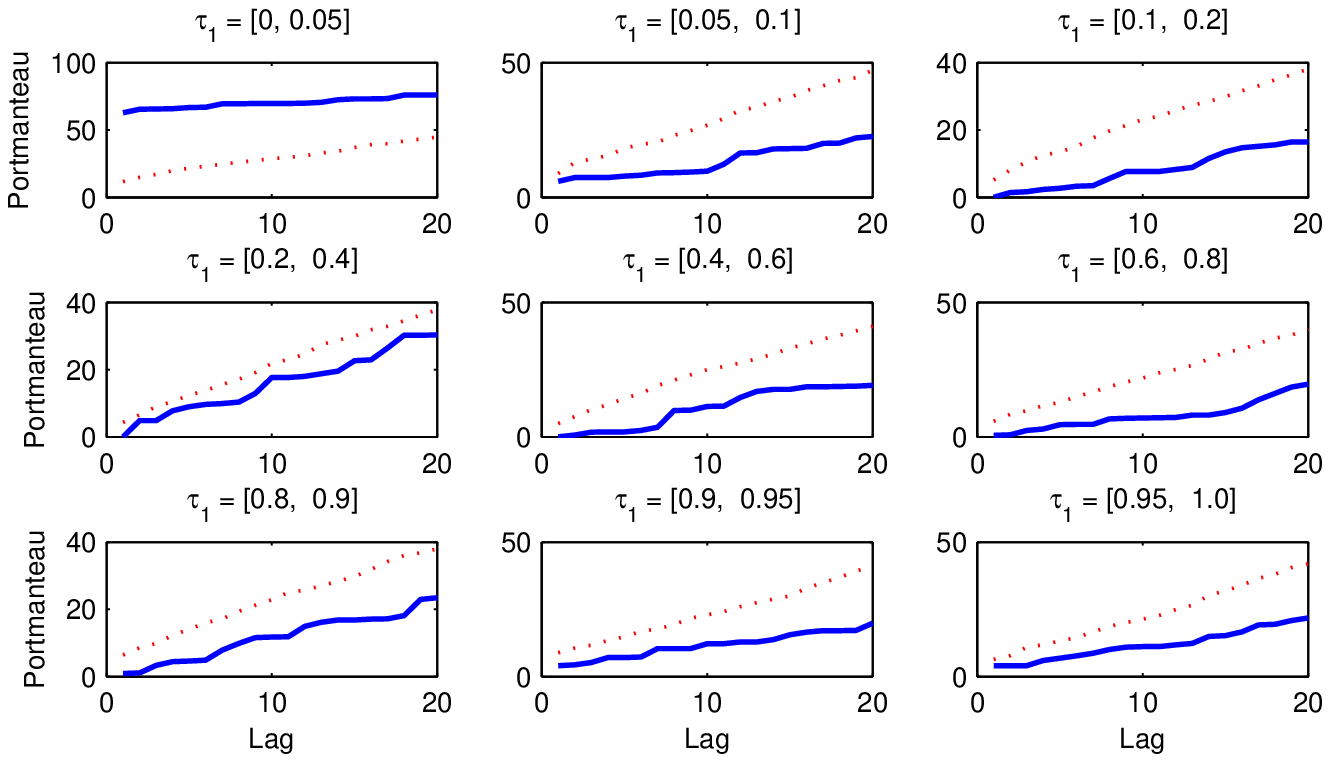';file-properties "XNPEU";}}
\end{center}

{\small \noindent Figure 9(b). [US to UK, std. residual] Box-Ljung test
statistic }${\small \hat{Q}}_{\tau }^{(p)}${\small \ for each lag }${\small p%
}${\small \ using }${\small \hat{\rho}}_{\tau }{\small (k)}${\small . Same
as Figure 1(b).\pagebreak\ }

\begin{center}
\FRAME{itbpF}{6.263in}{3.2499in}{0in}{}{}{sdgjr_tousfromuk.eps}{\special%
{language "Scientific Word";type "GRAPHIC";maintain-aspect-ratio
TRUE;display "PICT";valid_file "F";width 6.263in;height 3.2499in;depth
0in;original-width 6.3543in;original-height 3.2827in;cropleft "0";croptop
"1";cropright "1";cropbottom "0";filename
'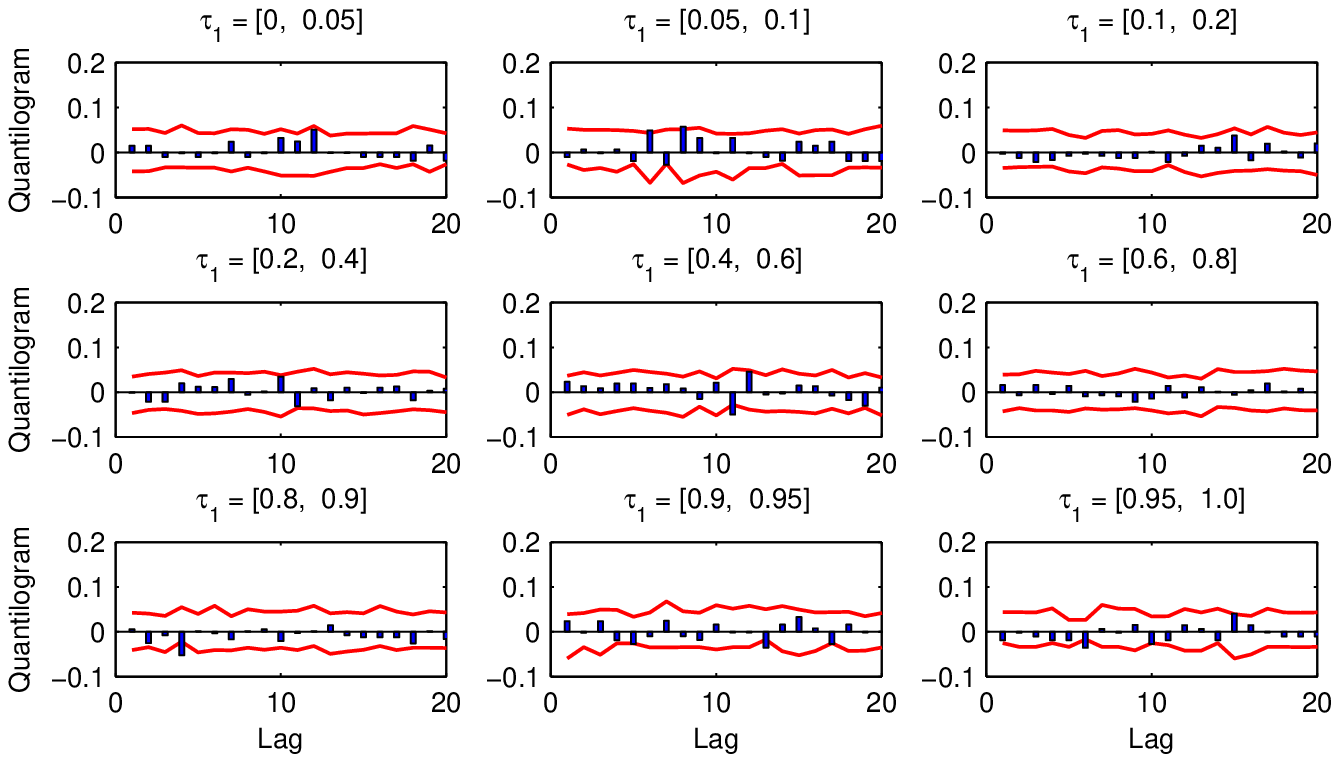';file-properties "XNPEU";}}
\end{center}

{\small \noindent Figure 10(a). [UK to US, std. residual] Cross-quantilogram 
}${\small \hat{\rho}}_{\tau }{\small (k)}${\small \ to detect directional
predictability from UK to US using the standardized residual from the
GJR-GARCH model. }$\tau _{1}${\small =}$\tau _{2}${\small . Same as Figure
1(a).}

\begin{center}
\FRAME{itbpF}{6.263in}{3.2499in}{0in}{}{}{sdgjr_tousfromuk_q.eps}{\special%
{language "Scientific Word";type "GRAPHIC";maintain-aspect-ratio
TRUE;display "PICT";valid_file "F";width 6.263in;height 3.2499in;depth
0in;original-width 6.3543in;original-height 3.2827in;cropleft "0";croptop
"1";cropright "1";cropbottom "0";filename
'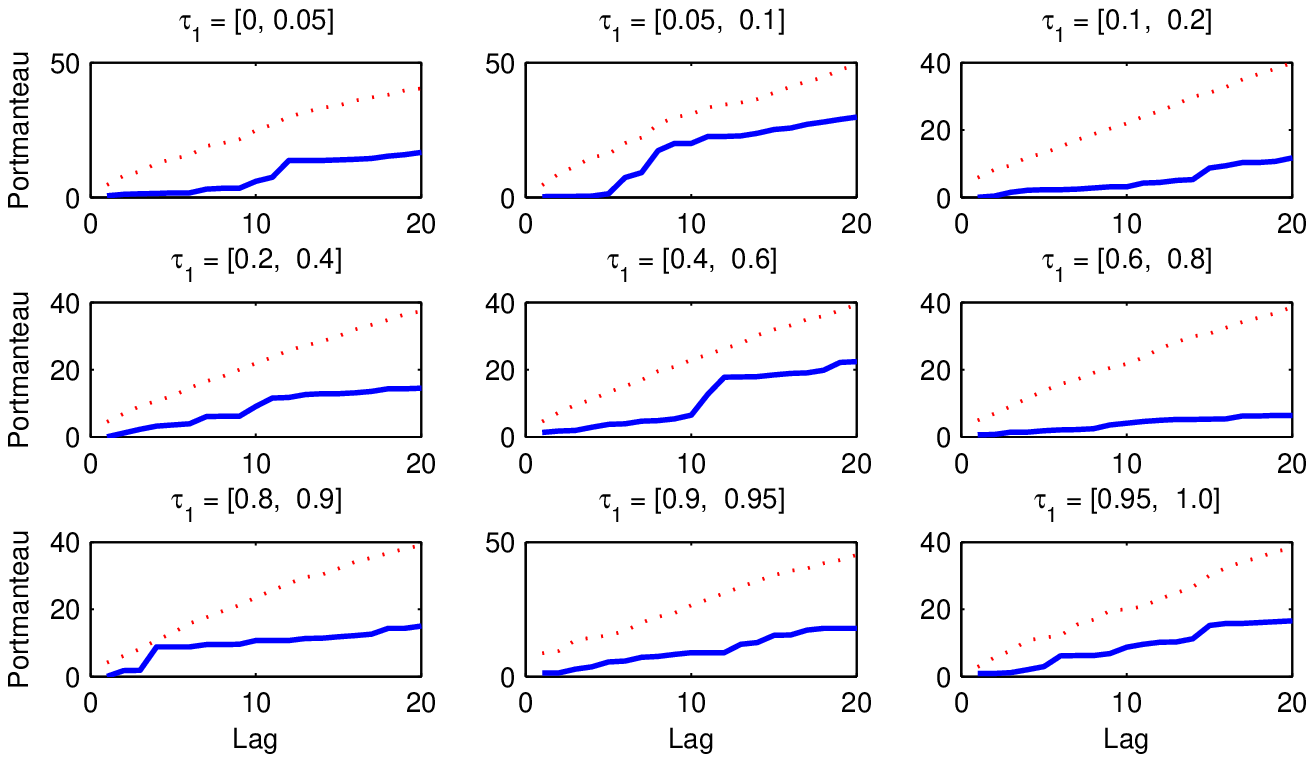';file-properties "XNPEU";}}
\end{center}

{\small \noindent Figure 10(b). [UK to US, std. residual] Box-Ljung test
statistic }${\small \hat{Q}}_{\tau }^{(p)}${\small \ for each lag }${\small p%
}${\small \ using }${\small \hat{\rho}}_{\tau }{\small (k)}${\small . Same
as Figure 1(b).\pagebreak\ }

\begin{center}
\FRAME{itbpF}{6.263in}{3.2499in}{0in}{}{}{ustojp.eps}{\special{language
"Scientific Word";type "GRAPHIC";maintain-aspect-ratio TRUE;display
"PICT";valid_file "F";width 6.263in;height 3.2499in;depth 0in;original-width
6.3543in;original-height 3.2827in;cropleft "0";croptop "1";cropright
"1";cropbottom "0";filename '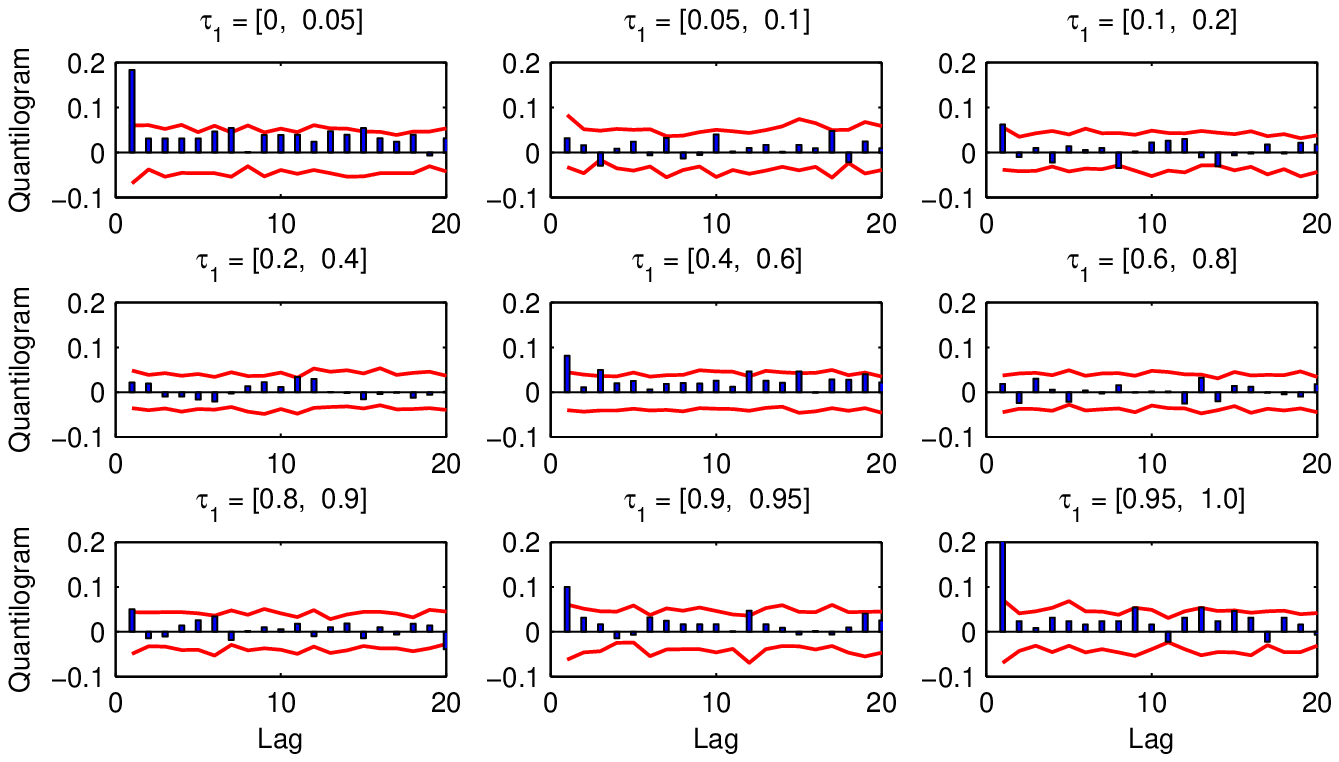';file-properties "XNPEU";}}
\end{center}

{\small \noindent Figure 11(a). [US to Japan] Cross-quantilogram }${\small 
\hat{\rho}}_{\tau }{\small (k)}${\small \ to detect directional
predictability from US to Japan. }$\tau _{1}${\small =}$\tau _{2}${\small .
Same as Figure 1(a).}

\begin{center}
\FRAME{itbpF}{6.263in}{3.2499in}{0in}{}{}{sdgjr_ustojp.eps}{\special%
{language "Scientific Word";type "GRAPHIC";maintain-aspect-ratio
TRUE;display "PICT";valid_file "F";width 6.263in;height 3.2499in;depth
0in;original-width 6.3543in;original-height 3.2827in;cropleft "0";croptop
"1";cropright "1";cropbottom "0";filename '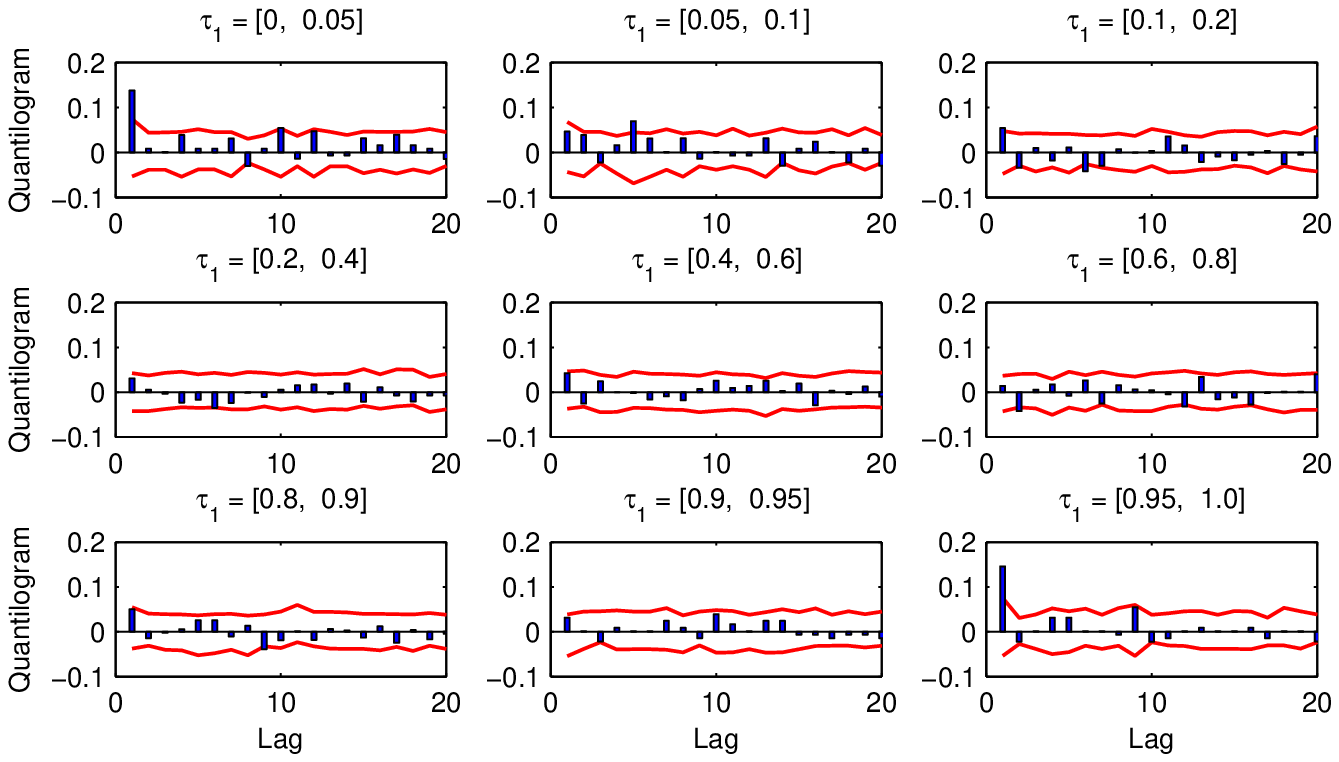';file-properties
"XNPEU";}}
\end{center}

{\small \noindent Figure 11(b). [US to Japan, std. residual]
Cross-quantilogram }${\small \hat{\rho}}_{\tau }{\small (k)}${\small \ to
detect directional predictability from US to Japan using the standardized
residual from the GJR-GARCH model. }$\tau _{1}${\small =}$\tau _{2}${\small %
. Same as Figure 1(a).}

{\small \pagebreak }

\begin{center}
\FRAME{itbpF}{6.263in}{3.2499in}{0in}{}{}{left_ustouk.eps}{\special{language
"Scientific Word";type "GRAPHIC";maintain-aspect-ratio TRUE;display
"PICT";valid_file "F";width 6.263in;height 3.2499in;depth 0in;original-width
6.3543in;original-height 3.2827in;cropleft "0";croptop "1";cropright
"1";cropbottom "0";filename '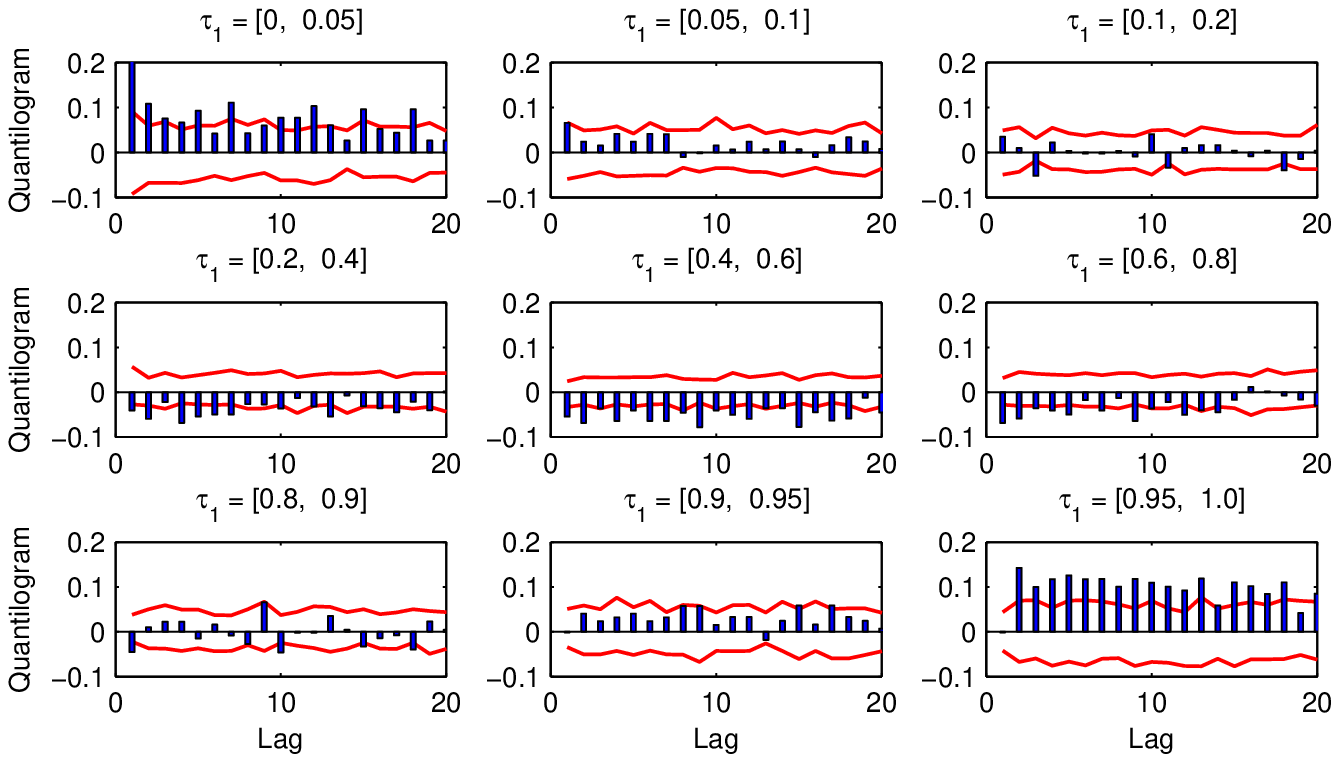';file-properties "XNPEU";}}
\end{center}

{\small \noindent Figure 12(a). [US to UK, from the left tail]
Cross-quantilogram }${\small \hat{\rho}}_{\tau }{\small (k)}${\small \ to
detect directional predictability from US to UK. }$\tau _{1}\neq \tau _{2}$%
{\small \ and }$\tau _{2}${\small =}${\small [0,0.05]}${\small \ where }$%
\tau _{2}${\small \ is the quantile range of US the stock return. Same as
Figure 1(a).}

\begin{center}
\FRAME{itbpF}{6.263in}{3.2499in}{0in}{}{}{right_ustouk.eps}{\special%
{language "Scientific Word";type "GRAPHIC";maintain-aspect-ratio
TRUE;display "PICT";valid_file "F";width 6.263in;height 3.2499in;depth
0in;original-width 6.3543in;original-height 3.2827in;cropleft "0";croptop
"1";cropright "1";cropbottom "0";filename '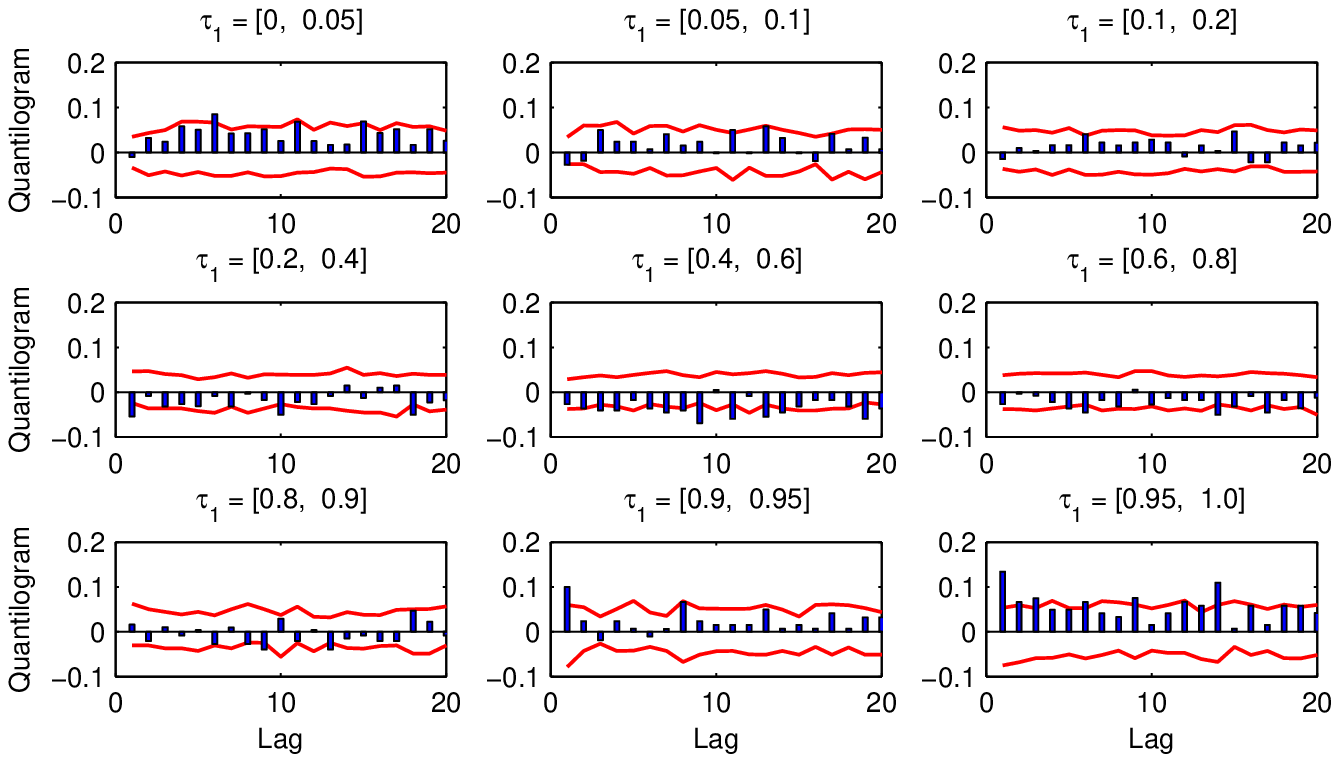';file-properties
"XNPEU";}}
\end{center}

{\small \noindent Figure 12(b). [US to UK, from the righttail]
Cross-quantilogram }${\small \hat{\rho}}_{\tau }{\small (k)}${\small \ to
detect directional predictability from US to UK. }$\tau _{1}\neq \tau _{2}$%
{\small \ and }$\tau _{2}${\small =}${\small [0.95,1]}${\small \ where }$%
\tau _{2}${\small \ is the quantile range of the US stock return. Same as
Figure 1(a).}

{\small \pagebreak\ }

\begin{center}
\FRAME{itbpF}{6.263in}{3.2499in}{0in}{}{}{left_ustouk_sdgjr.eps}{\special%
{language "Scientific Word";type "GRAPHIC";maintain-aspect-ratio
TRUE;display "PICT";valid_file "F";width 6.263in;height 3.2499in;depth
0in;original-width 6.3543in;original-height 3.2827in;cropleft "0";croptop
"1";cropright "1";cropbottom "0";filename
'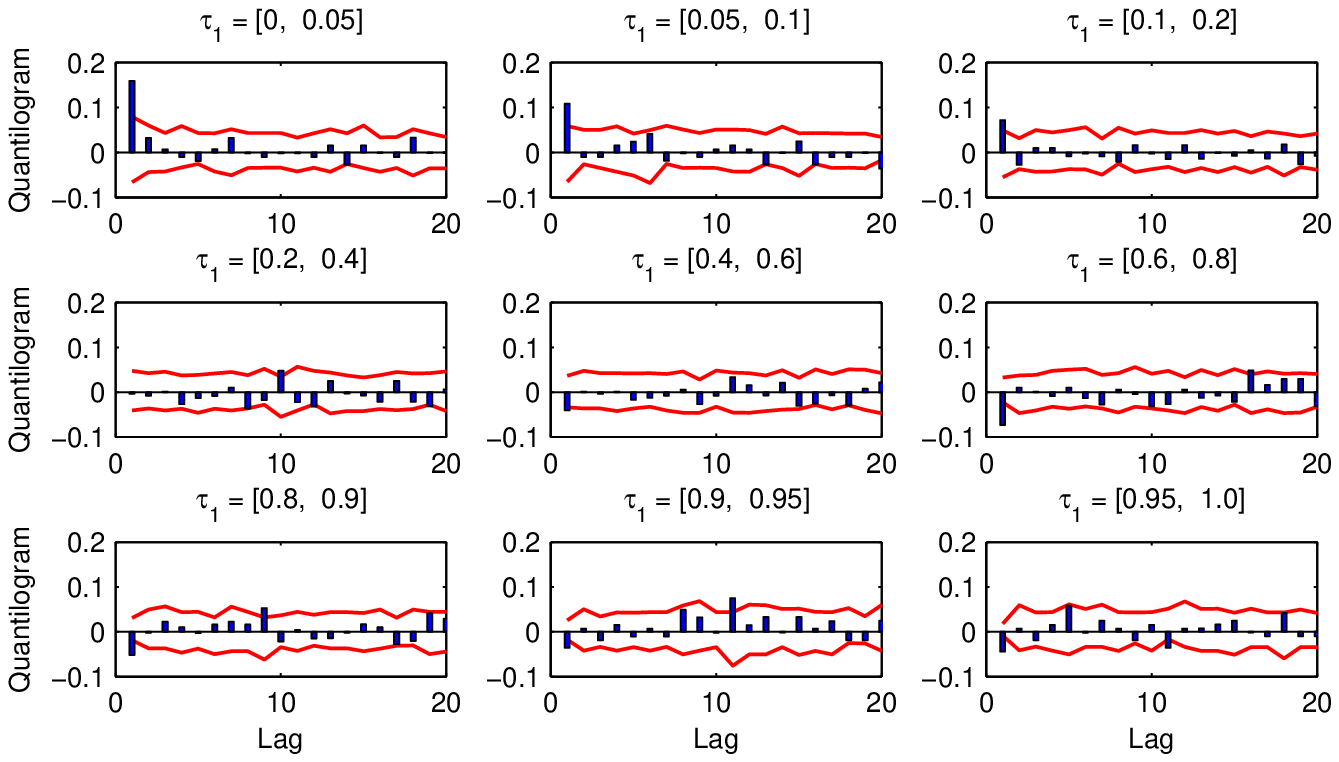';file-properties "XNPEU";}}
\end{center}

{\small \noindent Figure 13(a). [US to UK, std. residual, from the left
tail] Cross-quantilogram }${\small \hat{\rho}}_{\tau }{\small (k)}${\small \
from US to UK using the standardized residual from the GJR-GARCH model. }$%
{\small \tau }_{1}{\small \neq \tau }_{2}${\small \ and }${\small \tau }_{2}$%
{\small =}${\small [0,0.05]}${\small \ where }${\small \tau }_{2}${\small \
is for the US stock return. Same as Figure 1(a).}

\begin{center}
\FRAME{itbpF}{6.263in}{3.2499in}{0in}{}{}{right_ustouk_sdgjr.eps}{\special%
{language "Scientific Word";type "GRAPHIC";maintain-aspect-ratio
TRUE;display "PICT";valid_file "F";width 6.263in;height 3.2499in;depth
0in;original-width 6.3543in;original-height 3.2827in;cropleft "0";croptop
"1";cropright "1";cropbottom "0";filename
'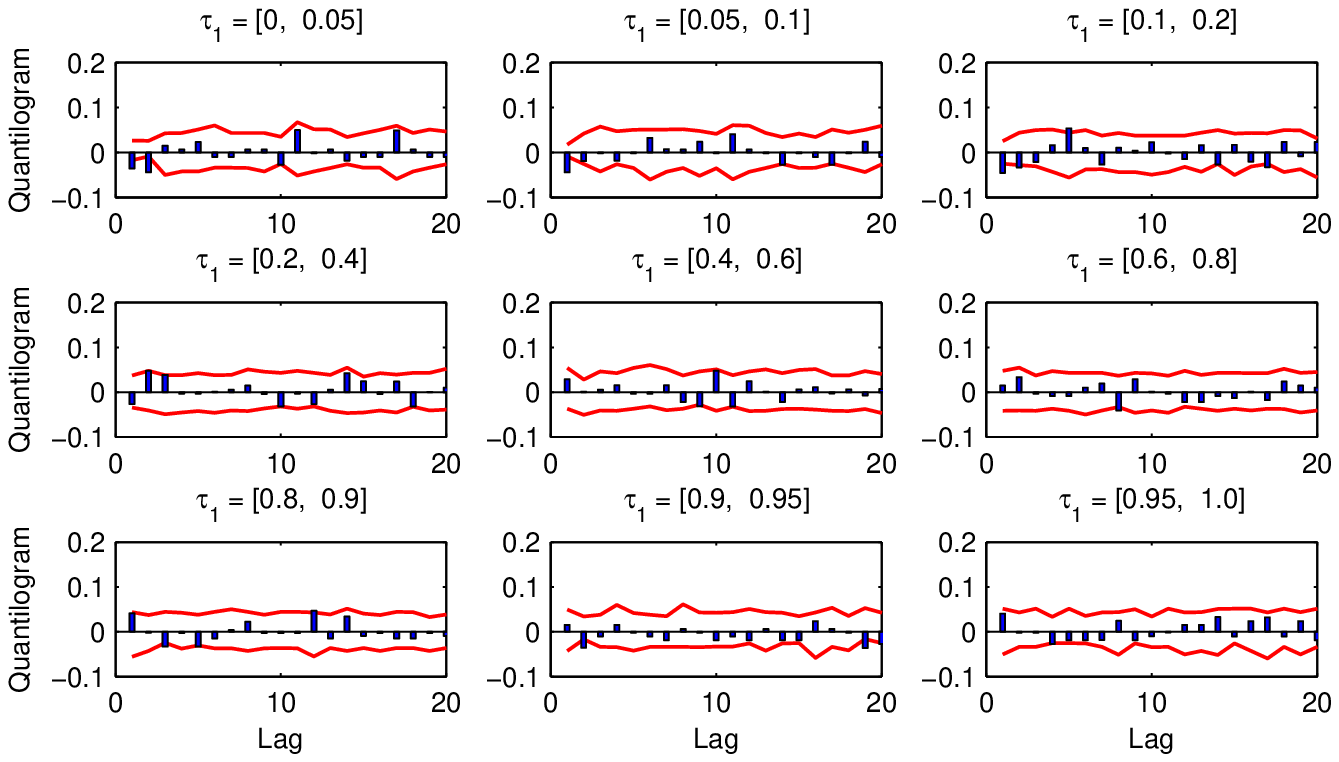';file-properties "XNPEU";}}
\end{center}

{\small \noindent Figure 13(b). [US to UK, std. residual, from the right
tail] Cross-quantilogram }${\small \hat{\rho}}_{\tau }{\small (k)}${\small \
from US to UK for }${\small \tau }_{1}{\small \neq \tau }_{2}${\small \ and }%
${\small \tau }_{2}${\small =}${\small [0.95,1].}${\small \ Same as Figure
13(a).}

{\small \pagebreak\ }

\doublespacing{\small \noindent }

\setstretch{1.00}

%

\end{document}